\documentclass[preprints,review,accept,moreauthors,pdftex]{Definitions/mdpi} 

\usepackage{amssymb}
\usepackage{mathrsfs}
\usepackage{slashed}
\usepackage{bm}

\setitemize{parsep=6pt,itemsep=0pt,leftmargin=*,labelsep=5.5mm}
\setenumerate{parsep=6pt,itemsep=0pt,leftmargin=*,labelsep=5.5mm}
\setlist[description]{itemsep=0mm}

\newcommand{\Tr}{\mathop{\mathrm{Tr}}}

\def\MeV{\,{\rm MeV}}
\def\GeV{\,{\rm GeV}}

\makeatletter      
                    \renewcommand{\maketag@@@}[1]{\hbox{\m@th\normalsize\normalfont#1}}%
                    \makeatother

\firstpage{1} 
\pubvolume{01}
\issuenum{01}
\articlenumber{00005}
\pubyear{2020}
\copyrightyear{2020}
\history{Received: date; Accepted: date; Published: date}

\pdfoutput=1

\Title{New Physics of Strong Interaction and Dark Universe}

\Author{Vitaly Beylin $^{1}$, Maxim Khlopov $^{1,2,3,}$*, Vladimir Kuksa $^{1}$ and Nikolay Volchanskiy $^{1,4}$}

\address{%
$^{1}$ \quad Institute of Physics, Southern Federal University, Stachki 194, 344090 Rostov on Don, Russia; beylinv@sfedu.ru (V.B.); vkuksa47@mail.ru (V.K.); nvolchanskiy@sfedu.ru (N.V.)\\
$^{2}$ \quad CNRS, Astroparticule et Cosmologie, Université de Paris, F-75013 Paris, France\\
$^{3}$ \quad National Research Nuclear University ``MEPHI'' (Moscow State Engineering Physics Institute), \mbox{31 Kashirskoe Chaussee,} \mbox{115409 Moscow, Russia}\\
$^{4}$ \quad Bogoliubov Laboratory of Theoretical Physics, Joint Institute for Nuclear Research, Joliot-Curie 6, \mbox{141980 Dubna, Russia}}

\corres{Correspondence: khlopov@apc.univ-paris7.fr; Tel.:+33-676380567}

\abstract{The history of dark universe physics can be traced from processes in the very early universe to the modern dominance of dark matter and energy. Here, we review the possible nontrivial role of strong interactions in cosmological effects of new physics. In the case of ordinary QCD interaction, the existence of new stable colored particles such as new stable quarks leads to new exotic forms of matter, some of which can be candidates for dark matter. New QCD-like strong interactions lead to new stable composite candidates bound by QCD-like confinement. We put special emphasis on the effects of interaction between new stable hadrons and ordinary matter, formation of anomalous forms of cosmic rays and exotic forms of matter, like stable fractionally charged particles. The possible correlation of these effects with high energy neutrino and cosmic ray signatures opens the way to study new physics of strong interactions by its indirect multi-messenger astrophysical probes.}


\begin{document}

\section{Introduction}
 The modern standard model of cosmology, involving inflation, baryosynthesis and dark matter/energy finds its basis beyond the standard model (BSM) of electroweak (EW) and strong interactions, thus moving the physics of the universe to the dark side of the fundamental physics. These phenomena determine the history of the cosmological evolution that resulted in the modern structure of the universe~\cite{Lindebook,Kolbbook,Rubakovbook1,Rubakovbook2,book,newBook,4,DMRev}.
 
 BSM physics not only provides a physical basis for the standard elements of cosmological construction, but~also inevitably involves new nonstandard cosmological and astrophysical features (see~\cite{ijmpd19,hadronic} for review and references). Here, we study such features, which may appear as signatures for the new physics of strong~interactions.
 
 In the strong interaction of the standard model (QCD), new physics comes from colored states with new quantum numbers. If~these new charges are conserved, the~lightest particle which possesses this property is stable and can have important cosmological impact. An~interesting feature of new stable heavy quarks is their binding by chromo-Coulomb forces in heavy quark clusters with strongly suppressed hadronic interaction, making their properties more close to the features of~leptons.
 
 BSM models can involve additional non-abelian symmetry, giving rise to new composite particles, whose constituents are bound by QCD-like confinement. Such states may have exotic features of multiple or fractional charge~leptons. 
 
 In the present review we discuss predictions of QCD and QCD-like models for possible new forms of stable matter and dark matter candidates (Section \ref{models}) as well as their effects and multi-messenger probes in the galaxy (Section \ref{galaxy}).

\section{New Physics from QCD and QCD-Like~Models}\label{models}\vspace{-6pt}
\subsection{General Features of New Physics of Strong Interactions in Dark~Cosmology}
Models adding new symmetry to the symmetry of the standard model (SM) predict new conserved quantum numbers, which provide stability of the lightest particle that possesses them. Such a particle can also possess QCD color and constitute new stable~hadrons. 

The addition of new non-abelian symmetry involves QCD-like interactions binding new QCD-like constituents in new composite states. Such composite states do not have ordinary hadronic interaction and look like leptons in their interaction with baryonic~matter.

These new stable particles can be dark matter (DM) candidates, since even hadronic interaction with cosmological plasma is not sufficiently strong to hinder the decoupling of gas of stable~hadrons.

\subsubsection{New Stable~Quarks}

Hadronic dark matter (HaDM) is one of the natural variants of strongly interacting dark matter scenario.
 The most popular candidates, weakly interacting massive particles (WIMP), have not yet been discovered. Moreover, strong restrictions on WIMP-nucleon scattering cross section~\cite{XENON18} exclude some variants of WIMP scenarios.  Some scenarios of DM
 with strongly interacting massive particles (SIMP) are presented in
Refs.~\cite{RanHuo16,DeLuca18,Beyl18,Beyl19,hadronic}. As~was noted in the introduction, usually strongly self-interacting dark matter scenarios are realized by introducing extra groups of gauge symmetry and additional sets of fields.
Here, we consider the scenario with hadronic DM, which contains a minimal set of new fields. Namely, heavy singlet quark. This scenario is realized in the framework of grand unification
scheme (for example, $E_6$-theory or $SU(5)$ supersymmetric extension contain $SU(2)$-singlet quark). We should note that hadronic DM is not only self-interacting; hadron-type DM
particles strongly interact with ordinary matter.
Here, we give a brief description of the origin and main properties of new stable quarks which enter the new heavy hadrons as
DM~particles.

In the scenario with heavy HaDM, new hadrons consist of new heavy stable quarks $Q$ and light standard ones, $q$. New quarks possess standard strong (QCD-type) interactions and, together
with standard quarks, form mesonic $M=(qQ)$ and fermionic $F_1=(qqQ), F_2=(qQQ), F_3=(QQQ)$ composite states. New heavy quarks arise in the extension with singlet quarks~\cite{Beyl18},
chiral-symmetric models 
~\cite{Beyl19}, \cite{Bazhutov17} and 4th generation standard model extensions \cite{Khlopov11,Khlopov14,Cudell15}. Principal properties of new heavy
hadrons and their phenomenology were presented in Refs.~\cite{Beyl18,Beyl19,hadronic}. It was underlined in these works that the repulsive character of DM-nucleon interactions makes it
possible to escape rigid cosmochemical restrictions on the relative concentration of anomalous hydrogen and helium~\cite{Bazhutov17,hadronic}. In~this subsection, we briefly present
a theoretical base of the HaDM scenario, which is constructed in the framework of the extension with a singlet quark~\cite{Beyl18} and chiral-symmetric model~\cite{Beyl19}.

The minimal Lagrangian of the extension of SM with new stable quarks is as follows:
\begin{equation}\label{1L}
L=L^{SM}+L^Q,
\end{equation}
where $L^Q$ describes interaction of new quarks $Q$ with gauge bosons. In~the case of singlet quark $Q_s$ Lagrangian is defined in standard way:
\begin{equation}\label{2L}
L^Q_s=i\bar{Q_s}\gamma^{\mu}(\partial_{\mu}-ig_1 q_Q V_{\mu} -ig_{st} t_a G^a_{\mu})Q_s - M_Q \bar{Q_s} Q_s.
\end{equation}

In (\ref{2L}), the~matrix $t_a=\lambda_a/2$ are generators of $SU_C(3)$ -group and $M_Q$ is the mass parameter of $Q$.

The chiral-symmetric extension of SM has an additional set of up and down quarks with anti-symmetric (with respect to standard one) chiral structure:
\begin{equation}\label{3L}
Q=\{Q_R=(U_R , D_R);
  \,\,\,U_L,\,\,\,D_L\}
\end{equation}

The structure of covariant derivatives follows from this definition:
\begin{align}\label{4L}
D_{\mu}Q_R=&(\partial_{\mu}-ig_1 Y_Q V_{\mu}-\frac{ig_2}{2} \tau_a V^a_{\mu}-ig_3 t_iG^i_{\mu}) Q_R;\notag\\
D_{\mu}U_L=&(\partial_{\mu}-ig_1 Y_U V_{\mu}-ig_3 t_iG^i_{\mu}) U_L,\notag\\D_{\mu}D_L=&(\partial_{\mu}-ig_1 Y_D V_{\mu}-ig_3 t_iG^i_{\mu}) D_L.
\end{align}

In the equations~in (\ref{4L}), the values $Y_A$ ($A=Q,U,D$) are hypercharges of quarks' doublets and singlets and $t_i$ are generators of the $SU_C(3)$-group. The~gauge fields $V^a_{\mu}$ are superheavy
chiral partners of standard gauge fields. Further, we consider some specific variant of the chiral-symmetric model with vector-like interaction of new quarks with gauge bosons
~\cite{hadronic}.

The structure of interactions in the extension with singlet quarks and in the chiral-symmetric model has a vector nature in both cases. So, we use universal expression for the Lagrangian model of EW interactions:
\begin{equation}\label{5L}
L(Q_a ,A,Z)=g_a(c_w A_{\mu}-s_w Z_{\mu})\bar{Q}_a\gamma^{\mu}Q_a,\,\,\,Q_a=Q_s ,U,D,
\end{equation}
where $g_a=g_1 q_a$ and $q_a=2/3$ or $q_a=-1/3$ for the case of up or down new quark. The~definitions of charges for quarks $U$ and $D$ in chiral-symmetrical scenario are standard
also. These assumptions make it possible to form neutrally coupled states with standard quarks. In~the models under consideration, vector interaction of new heavy quarks with gauge
vector fields gives small contributions to polarizations. So, the~contributions of new quarks into Peskin--Takeuchi (PT) parameters $S, T, U$, which describe their effects in
electroweak physics, are small too. Using standard definitions of the PT parameters and vertices in (\ref{5L}), by~a straightforward calculation, we get for the parameters $S$ and $U$
($T=0$) the following expressions:
\begin{equation}\label{6L}
S=-U=\frac{k s^4_w}{9\pi}[-\frac{1}{3}+2(1+2\frac{M^2_Q}{M^2_Z})(1-\sqrt{\beta}\arctan\frac{1}{\sqrt{\beta}})].
\end{equation}

Here, $\beta=4M^2_Q/M^2_Z -1$, $k=16(4)$ for the case of singlet quark model with the charge $q=2/3(-1/3)$, and~$k=20$ for the case of chiral-symmetric model. From~Equation~(\ref{6L}), it
follows that for heavy new quarks with mass $M_Q>500$ GeV the value of parameters $S=-U <10^{-2}$ is significantly
 less than the experimental limits~\cite{PDG18}:
\begin{equation}\label{7L}
S=0.00 +0.11(-0.10),\,\,\,U=0.08\pm 0.11,\,\,\,T=0.02+0.11(-0.12).
\end{equation}

 Thus, the~scenarios with vector-like new heavy quarks are not excluded by EW experimental restrictions.
There are, also, EW restrictions which follow from the presence of flavor-changing neutral currents (FCNC). In~the scenario under consideration, new quarks do not mix with standard
quarks and FCNC at the tree level are absent. So, there are no additional restrictions which follow from the rare processes, such as rare decays of mesons and the mixing in the systems
of neutral mesons~(oscillations).

The potential of interaction of new heavy mesons and nucleons, as~was shown in Ref.~\cite{hadronic}, has~repulsive character. So, new hadrons do not form coupled states with ordinary
matter and this effect excludes the formation of anomalous hydrogen and helium. Thus the scenario with hadronic DM does not contradict rigid cosmochemical restrictions on
the relative concentration of these elements~\cite{hadronic}.

\subsubsection{QCD-Like~Models}\label{sec212}

For almost forty years, many efforts have been invested in developing SM extensions conjecturing the existence of new nonperturbative BSM physics. In~particular, such models can postulate that the Higgs boson is a composite object consisting of some new fundamental constituents held together by an analog of strong force. There are many brilliant reviews surveying different aspects of this huge field in detail, e.g.,~\cite{Cacciapaglia:2020kgq, Panico:2015jxa, Csaki:2015hcd, DeGrand:2015zxa, Bellazzini:2014yua, Hill:2002ap, Schmaltz:2005ky, Chen:2006dy, Perelstein:2005ka}.

In this section, we consider some particular variants of models that extend SM by introducing an additional strong sector with heavy vector-like fermions, hyperquarks (H-quarks), charged under an H-color gauge group~\cite{Sundrum,Pasechnik:2013bxa,Pasechnik:2013kya,Lebiedowicz:2013fta,Pasechnik:2014ida,doi:10.1142/S0217751X17500427,2015JHEP...12..031A,2017PhRvD..95c5019A,2018JHEP...08..017B,2015JHEP...01..157A,2017JHEP...10..210M,Appelquist_ref}. Depending on H-quark quantum numbers, such models can encompass scenarios with composite Higgs doublets (see, e.g.,\ \cite{Cai:2018tet}) or a small mixing between fundamental Higgs fields of SM and composite hadron-like states of new strong sector making the Higgs boson partially composite. Models of this class leave room for the existence of DM candidates whose decays are forbidden by accidental symmetries. Besides, H-color models comply well with electroweak  precision constraints, since H-quarks are assumed to be~vector-like.

Among the simplest realizations of the scenario described are models with two or three vector-like H-flavors confined by strong H-color force $\text{Sp}(2\chi_{\tilde{c}})$, $\chi_{\tilde{c}} \geqslant 1$. The~models with H-color group SU(2)~\cite{2017PhRvD..95c5019A,Beylin:2016kga} are included as particular cases in this consideration due to isomorphism $\text{SU}(2) = \text{Sp}(2)$ \cite{2017PhRvD..95c5019A,Beylin:2016kga}. The~global symmetry group of the strong sector with symplectic H-color group is larger than for the special unitary case---it is the group SU($2n_F$) broken spontaneously to Sp($2n_F$), with~$n_F$ being a number of H-flavors. Going beyond the simplest (two-flavor) model is of interest because the phenomenology of such models is richer involving new fractionally charged states that can be stable. We posit that the extensions of SM under consideration preserve the elementary Higgs doublet in the set of Lagrangian field operators. This doublet mixes with H-hadrons, which makes the physical Higgs partially composite. The same coset SU($2n_F$)/Sp($2n_F$) can be used to construct composite two Higgs doublet model~\cite{Cai:2018tet} or little Higgs models~\cite{Low:2002ws,Csaki:2003si,Gregoire:2003kr,Han:2005dz,Brown:2010ke,Gopalakrishna:2015dkt}.

It should be also noted that there are multiple options of assigning electroweak quantum numbers to new H-quarks charged under symplectic color gauge group. In the two-flavor case, for example, there are two possibilities. Except for the model with vector-like H-quarks considered in this paper, one can also build a model with one left-handed doublet and two right-handed quark singlets (e.g., see \cite{Cacciapaglia:2020kgq}). 

Let us consider a model with the symmetry $G=G_\text{SM} \times \text{Sp}(2\chi_{\tilde{c}})$, $\chi_{\tilde{c}} \geqslant 1$, with~$G_\text{SM}$ and  $\text{Sp}(2\chi_{\tilde{c}})$ being the SM gauge group and a symplectic hypercolor group respectively. In~its field content, the~model introduces a doublet and a singlet of heavy vector-like H-quarks charged under H-color group. Then, in~the renormalizable case, the~most general Lagrangian invariant under $G$ reads
\begin{gather}\label{eq:LQS1}
    \mathscr{L} = \mathscr{L}_\text{SM} - \frac14 H^{\mu\nu}_{\underline{a}} H_{\mu\nu}^{\underline{a}} + i \bar Q \slashed{D} Q - m_Q \bar{Q} Q + i \bar S \slashed{D} S - m_S \bar{S} S + \delta\mathscr{L}_{\text{Y}},
\\ \label{eq:DQDS}
    D^\mu Q = \left[ \partial^\mu + \frac{i}{2} g_1 Y_Q B^\mu - \frac{i}{2} g_2 W_a^\mu \tau_a - \frac{i}2 g_{\tilde{c}} H^\mu_{\underline{a}} \lambda_{\underline{a}} \right] Q,
\qquad
    D^\mu S = \left[ \partial^\mu + i g_1 Y_S B^\mu - \frac{i}2 g_{\tilde{c}} H^\mu_{\underline{a}} \lambda_{\underline{a}} \right] S,
\end{gather}
where $H^\mu_{\underline{a}}$, $\underline{a} = 1 \dots \chi_{\tilde{c}} (2\chi_{\tilde{c}} + 1)$ are hypergluon fields and $H^{\mu\nu}_{\underline{a}}$ are their strength tensors; $\tau_a$ are the Pauli matrices; $\lambda_{\underline a}$, $\underline{a} = 1 \dots \chi_{\tilde{c}} (2\chi_{\tilde{c}} + 1)$ are $\text{Sp}(2\chi_{\tilde{c}})$ generators satisfying the relation
\begin{align}\label{eq:scgr}
\lambda_{\underline a}^\text{T} \omega + \omega \lambda_{\underline a} = 0,
\end{align}
where $\text{T}$ stands for ``transpose'', $\omega$ is an antisymmetric $2\chi_{\tilde{c}} \times 2\chi_{\tilde{c}}$ matrix, $\omega^\text{T} \omega = 1$. Hereafter, all~underscored indices correspond to representations of the H-color group $\text{Sp}(2\chi_{\tilde{c}})$.~In~the Lagrangian~\eqref{eq:LQS1}, the~contact Yukawa couplings $\delta\mathscr{L}_{\text{Y}}$ of the H-quarks and the SM Higgs doublet $\mathscr{H}$ are permitted by the symmetry $G$ if the hypercharges $Y_Q$ and $Y_S$ satisfy an additional linear relation: 
\begin{gather}\label{eq:LQHS}
    \delta\mathscr{L}_{\text{Y}} = y_\text{L} \left( \bar{Q}_\text{L} \mathscr{H} \right) S_\text{R} + y_\text{R} \left( \bar{Q}_\text{R} \varepsilon \bar{\mathscr{H}} \right) S_\text{L} + \text{h.c.} \quad \text{ for } \frac{Y_Q}{2}-Y_S = +\frac12;
\\
    \delta\mathscr{L}_{\text{Y}} = y_\text{L} \left( \bar{Q}_\text{L} \varepsilon \bar{\mathscr{H}} \right) S_\text{R} + y_\text{R} \left( \bar{Q}_\text{R} \mathscr{H} \right) S_\text{L} + \text{h.c.} \quad \text{ for } \frac{Y_Q}{2}-Y_S = -\frac12.
\end{gather}

It is a simple exercise to prove that the hypercolor part of the H-quark Lagrangian \eqref{eq:LQS1} can be rewritten in terms of a left-handed sextet as follows: 
\begin{gather}\label{eq:LP}
    \delta \mathscr{L}_\text{H-quarks, kin} = i \bar P_\text{L} \slashed{D} P_\text{L} ,
    \qquad P_\text{L} = \begin{pmatrix} Q_{\text{L}} \\ \epsilon \omega  Q_{\text{R}}{}^\text{C} \\ S_{\text{L}} \\ - \omega S_{\text{R}}{}^\text{C} \end{pmatrix},     \qquad
    D^\mu P_\text{L} = \left[ \partial^\mu - \frac{i}2 g_{\tilde{c}} H^\mu_{\underline{a}} \lambda_{\underline{a}} \right] P_\text{L},
\end{gather}
where $\epsilon = i \tau_2$, the~operation C denotes the charge conjugation.
Equation~\eqref{eq:LP} makes it obvious that, in~the absence of the electroweak interactions, the~H-quark Lagrangian is invariant under an extension of the chiral symmetry---a global SU(6) symmetry, dubbed the Pauli--G\"{u}rsey  symmetry \cite{Pauli1957Conservation, Gursey1958Relation}. The~subgroups of the SU(6) symmetry~include:  
\begin{itemize}
\item The chiral symmetry $\text{SU}(3)_\text{L} \times \text{SU}(3)_\text{R}$;
\item SU(4) subgroup corresponding to the two-flavor model without singlet H-quark $S$;
\item Two-flavor chiral group $\text{SU}(2)_\text{L} \times \text{SU}(2)_\text{R}$, which is a subgroup of both former subgroups.  
\end{itemize}

The global symmetry is broken both explicitly and~dynamically:  
\begin{itemize}
\item Explicitly---by the electroweak and Yukawa interactions, \eqref{eq:DQDS} and \eqref{eq:LQHS}, and~the H-quark masses;  
\item Dynamically---by H-quark condensate~\cite{Vysotskii1985Spontaneous,Verbaarschot2004Supersymmetric}:
\begin{gather}\label{eq:LTQ}
    \langle \bar QQ + \bar SS \rangle = \frac12 \langle  \bar P_\text{L} M_0 P_\text{R} + \bar P_\text{R} M_0^\dagger P_\text{L} \rangle,
\qquad P_\text{R}  = \omega P_\text{L}{}^\text{C},
    \qquad  M_0 = \begin{pmatrix} 0 & \varepsilon & 0\\ \varepsilon & 0 & 0 \\ 0 & 0 & \varepsilon \end{pmatrix}.
\end{gather}
\end{itemize}

The condensate \eqref{eq:LTQ} is invariant under $\text{Sp}(6) \subset \text{SU}(6)$ transformations $U$ that satisfy a condition
\begin{gather}\label{eq:symprel}
    U^\text{T} M_0 + M_0 U =0,
\end{gather}
i.e., the global SU(6) symmetry is broken dynamically to its Sp(6) subgroup. The~mass terms of H-quarks in \eqref{eq:LQS1} could break the symmetry further to  $\text{Sp(4)}\times\text{Sp}(2)$:
\begin{gather}\label{eq:massterm}
    \delta \mathscr{L}_\text{H-quarks, masses} = -\frac12 \bar P_\text{L} M_0' P_\text{R} + \text{h.c.},
\qquad M'_0 = -M'_0{}^\text{T} = \begin{pmatrix} 0 & m_Q \varepsilon & 0\\ m_Q \varepsilon & 0 & 0 \\ 0 & 0 & m_S \varepsilon \end{pmatrix}.
\end{gather}

It should be noted that the model under consideration is free of the gauge anomalies and can be easily reconciled with the electroweak precision constraints, since the H-quarks are vector-like, i.e.,~their electroweak interactions are chirally~symmetric.

The effective interactions of H-hadrons can be described in a linear $\sigma$-model involving the fundamental (not composite) Higgs doublet $\mathscr{H}$ and constituent H-quarks as independent degrees of freedom. The~Lagrangian of the model can be broken down into four sectors---(1) a sector of the constituent H-quarks (containing interactions of the quarks with gauge bosons), (2) Yukawa interactions of the (pseudo)scalars with the H-quarks, (3) a sector of (pseudo)scalar fields (which~produces their self-interactions and interactions with the Higgs boson), and~(4) terms communicating the explicit breaking of the SU($2n_F$) global symmetry to the effective fields:\footnote{Although the non-invariant terms responsible for the explicit symmetry breaking can be chosen in a variety of ways~\cite{1979PhRvC..19.1965C,2000PhRvC..61b5205D,1996NuPhA.603..239D}, we constrain ourselves to the most obvious tadpole-like one (as in~\cite{1969RvMP...41..531G,2013PhRvD..87a4011P}, for~example).}\vspace{6pt}
\begin{gather}\label{eq:ctq}
    \mathscr{L}_{\text{L}\sigma} = \mathscr{L}_\text{H-quarks} + \mathscr{L}_\text{Y} + \mathscr{L}_\text{scalars} + \mathscr{L}_\text{SB},
\\[3mm]
    \mathscr{L}_\text{H-quarks}  = i \bar P_\text{L} \slashed{D} P_\text{L},
\qquad
    \mathscr{L}_\text{Y} =  -\sqrt2 \varkappa \left( \bar P_\text{L} M P_\text{R} + \bar P_\text{R} M^\dagger P_\text{L} \right),
\\ \label{eq:L-scalars}
    \mathscr{L}_\text{scalars} =  D_\mu \mathscr{H}^\dagger \cdot D^\mu \mathscr{H} + \Tr D_\mu M^\dagger \cdot D^\mu M - U_\text{scalars},
\qquad
\mathscr{L}_\text{SB} = -\zeta \langle \bar Q Q + \bar S S \rangle (u+\sigma').
\end{gather}

Here, $\varkappa$ is a coupling constant; the parameter $\zeta$ is proportional to the current mass $m_Q$ of the H-quarks (see~\cite{1969RvMP...41..531G,2013PhRvD..87a4011P}, for~example); $M$ is a complex antisymmetric $2n_F \times 2n_F$ matrix of (pseudo)scalar fields whose singlet component develops a v.e.v. $u \sim -\langle \Tr \left( M M_0\right) \rangle$; the multiplets $P_\text{L, R}$ correspond now to the constituent H-quarks that are postulated not to interact with H-gluons but interact with the intermediate gauge bosons in the same way as the fundamental H-quarks. The~fields transform under the global symmetry SU($2n_F$) as follows:
\begin{gather}\label{eq:M_transf}
    M \to UMU^\text{T}, \qquad P_\text{L} \to U P_\text{L}, \qquad P_\text{R} \to \bar{U} P_\text{R}, \qquad U \in \text{SU}(2n_F),
\end{gather}
where $\bar{U}$ is the complex conjugate of $U$. These transformation laws allow one to define the covariant derivatives of the fields $P_\text{L, R}$ and $M$ easily (see explicit expressions in~\cite{hadronic}).

The physical (pseudo)scalar components of the field $M$ are listed in Table~\ref{tab:H-hadrons}. They include heavier analogs of all the light mesons of 3-flavor QCD and a set of H-baryons (H-diquarks)---singlets $A$ and $B$, and~doublets $\mathscr{A}$ and $\mathscr{B}$. The~Lagrangian of H-quark--H-hadron interactions reads
\begin{small}
\begin{gather}
    \mathscr{L}_\text{H-quarks} + \mathscr{L}_\text{Y} = i \bar Q \slashed{D} Q + i \bar S \slashed{D} S - \varkappa u \left( \bar Q Q + \bar S S \right)
    \notag\\
    -\varkappa \bar{Q} \left[ \sigma' + \frac{1}{\sqrt3} f + i \left( \eta + \frac{1}{\sqrt3} \eta' \right) \gamma_5
     + \left( a_a + i \pi_a \gamma_5 \right) \tau_a
    \right]  Q
    -\varkappa \bar{S} \left[ \sigma' - \frac{2}{\sqrt3} f
 + i \left( \eta - \frac{2}{\sqrt3} \eta' \right) \gamma_5 \right] S
   \notag\\
    -\sqrt2 \varkappa \left[ \left( \bar Q \mathscr{K}^\star \right) S + i \left( \bar Q \mathscr{K} \right) \gamma_5 S + \text{h.c.} \right]
	 -\sqrt2 \varkappa \left[ \left( \bar Q \mathscr{A} \right) \omega S^\text{C}
    + i \left( \bar Q \mathscr{B} \right) \gamma_5 \omega S^\text{C} + \text{h.c.} \right]
   \notag\\ 
    -\frac{\varkappa}{\sqrt2}  \left( A \bar Q \varepsilon \omega Q^\text{C}
    +iB \bar Q \gamma_5 \varepsilon \omega Q^\text{C} + \text{h.c.} \right) , \label{eq:LSM-lagr-QS-phys}
\end{gather}
\end{small}
\begin{gather}\label{eq:QS-cd}
    D_\mu Q = \partial_\mu Q + \frac{i}2 g_1 Y_Q B_\mu Q - \frac{i}2 g_2 W_\mu^a \tau_a Q,
    \qquad
    D_\mu S = \partial_\mu S + i g_1 Y_S B_\mu S,
\end{gather}
where $\mathscr{K}^\star$, $\mathscr{K}$ and $\mathscr{A}$, $\mathscr{B}$ are $\text{SU}(2)_\text{L}$ doublets of H-mesons and H-baryons respectively. The~Lagrangian for the case of two-flavor model, $n_F = 2$, is obtained by simply neglecting all terms with the singlet H-quark $S$ in Equation~\eqref{eq:LSM-lagr-QS-phys}.

The kinetic terms of the (pseudo)scalars in the Lagrangian \eqref{eq:L-scalars} produce interactions of the H-hadrons with the gauge bosons:
\begin{align}\label{eq:TM}
    \mathscr{T}_\text{scalars}  = \frac12 \sum_\varphi D_\mu \varphi \cdot D^\mu \varphi  + \sum_\Phi \left( D_\mu \Phi \right)^\dagger D^\mu \Phi
	 + D_\mu \bar A \cdot D^\mu A + D_\mu \bar B \cdot D^\mu B ,
\end{align}
where $\varphi = h$, $h_a$, $\pi_a$, $a_a$, $\sigma$, $f$, $\eta$, $\eta'$ are singlet and triplet fields, $\Phi = \mathscr{K}$, $\mathscr{K}^\star$, $\mathscr{A}$, $\mathscr{B}$ are doublets. The~fields $h$ and $h_a$, $a=1$, 2, 3 are components of the fundamental Higgs doublet $\mathscr{H} = \frac{1}{\sqrt2} (h+i h_a\tau_a) \left( \begin{smallmatrix} 0 \\ 1 \end{smallmatrix} \right)$. The~covariant derivatives in the Lagrangian \eqref{eq:TM} are defined as follows:\vspace{6pt}
\begin{gather}\label{eq:dpi}
	D_\mu h
	= \partial_\mu h
             +\frac12 (g_1 \delta_3^a B_\mu +g_2 W_\mu^a ) h_a,
	\qquad\qquad
	D_\mu \phi
	= \partial_\mu \phi, \quad \phi=\sigma, \, f, \, \eta, \, \eta',
	\\
    D_\mu h_a
        =\partial_\mu h_a
             -\frac12 (g_1 \delta_3^a B_\mu +g_2 W_\mu^a ) h
             -\frac12 e_{abc} (g_1 \delta_3^b B_\mu -g_2 W_\mu^b ) h_c,
	\\
    D_\mu M_a = \partial_\mu M_a + g_2 e_{abc} W_\mu^b M_c, \quad M = \pi,\, a,
	\qquad\qquad
	D_\mu Z = \partial_\mu Z + i g_1 Y_Q B_\mu Z, \quad Z = A,\, B,
    \\
    D_\mu \mathscr{K} = \left[ \partial_\mu + i g_1 \left( \frac{Y_Q}{2}-Y_S \right ) B_\mu - \frac{i}{2} g_2 W_\mu^a \tau^a \right] \mathscr{K},
	\\
	D_\mu \mathscr{K}^\star = D_\mu \mathscr{K} \bigr|_{\mathscr{K} \to \mathscr{K}^\star},
	\quad
	D_\mu \mathscr{A} = D_\mu \mathscr{K} \biggr|{}_{ \begin{subarray}{l} \mathscr{K} \to \mathscr{A} \\ Y_S \to -Y_S \end{subarray} },
	\quad
	D_\mu \mathscr{B} = D_\mu \mathscr{K} \biggr|{}_{ \begin{subarray}{l} \mathscr{K} \to \mathscr{B} \\ Y_S \to -Y_S \end{subarray} }.
\end{gather}

\begin{table}[H]
\caption{The lightest (pseudo)scalar H-hadrons in $\text{Sp}(2\chi_{\tilde{c}})$ model with two and three flavors of H-quarks (in the limit of vanishing mixings). The~lower half of the table lists the states present only in the three-flavor version of the model containing the singlet H-quark $S$. $T$ is the weak isospin. $\tilde G$ denotes hyper-$G$-parity of a state. $\tilde B$ is the H-baryon number. $Q_\text{em}$ is the electric charge (in units of the positron charge $e=|e|$). The~H-quark charges are $Q^U_\text{em}  = (Y_Q+1)/2$, $Q^D_\text{em} = (Y_Q-1)/2$, and~$Q^S_\text{em} = Y_S$, which~is seen from \eqref{eq:QS-cd}.}
\centering
{\begin{tabular}{ccccccccc}
\toprule
\textbf{State} & $$ & \textbf{H-Quark Current} & $$ & $\bm{T^{\tilde G}(J^{PC})}$ &  & $\bm{\tilde{B}}$ &  & $\bm{Q_\text{\textbf{em}}}$ \\
\midrule
$\sigma$ &  & $\bar Q Q + \bar SS$ &  & $0^+(0^{++})$ & $$ & 0 & $$ & 0 \\
$\eta$ &  & $i \left( \bar Q \gamma_5 Q + \bar S \gamma_5 S \right)$ &  & $0^+(0^{-+})$ & $$ & 0 & $$ & 0 \\
$ a_k$ &  & $\bar Q \tau_k Q$ &  & $1^-(0^{++})$ & $$ & 0 & $$ & $\pm 1$, 0 \\
$\pi_k$ &  & $i \bar Q \gamma_5 \tau_k Q$ &  & $1^-(0^{-+})$ & $$ & 0 & $$ & $\pm 1$, 0 \\
$A$ &  & $\bar Q^\text{C} \varepsilon \omega Q$ &  & $0^{\hphantom{+}}(0^{-\hphantom{+}})$ &  & 1 &  & $Y_Q$ \\
$B$ &  & $i \bar Q^\text{C}  \varepsilon \omega \gamma_5 Q$ &  & $0^{\hphantom{+}}(0^{+\hphantom{+}})$ & $$ & 1 & $$ & $Y_Q$ \\
\midrule
$f$ &  & $\bar Q Q -2 \bar SS$ & & $0^+(0^{++})$ & $$ & 0 & $$ & 0 \\
$\eta'$ &  & $i \left( \bar Q \gamma_5 Q - 2 \bar S \gamma_5 S \right)$ &  & $0^+(0^{-+})$ & $$ & 0 & $$ & 0 \\
$\mathscr{K}^\star$ &  & $\bar S Q$ &  & $\frac12^{\hphantom{+}}(0^{+\hphantom{+}})$ & $$ & 0 & $$ & $Y_Q/2-Y_S \pm 1/2$ \\
$\mathscr{K}$ &  & $i \bar S \gamma_5 Q$ &  & $\frac12^{\hphantom{+}}(0^{-\hphantom{+}})$ & $$ & 0 & $$ & $Y_Q/2-Y_S \pm 1/2$ \\
$\mathscr{A}$ &  & $\bar S^\text{C} \omega Q$ &  & $\frac12^{\hphantom{+}}(0^{-\hphantom{+}})$ & $$ & 1 & $$ & $Y_Q/2+Y_S \pm 1/2$ \\
$\mathscr{B}$ &  & $i \bar S^\text{C} \omega \gamma_5 Q$ &  & $\frac12^{\hphantom{+}}(0^{+\hphantom{+}})$ & $$ & 1 & $$ & $Y_Q/2+Y_S \pm 1/2$ \\
\bottomrule
\end{tabular} \label{tab:H-hadrons} }
\end{table}

For simplicity, we consider only renormalizable interactions of the scalar fields---the Higgs boson and (pseudo)scalar H-hadrons. Therefore, the~corresponding potential can be written as follows:
\begin{gather}\label{eq:U}
    U_\text{scalars} = \sum_{i=0}^4 \lambda_i I_i + \sum_{0 = i \leqslant k = 0}^3 \lambda_{ik} I_i I_k.
\end{gather}

Here, $I_i$, $i=0$, 1, 2, 3, 4 are the lowest dimension invariants
\begin{small}
\begin{gather}\label{eq:Uinvs}
    I_0 = \mathscr{H}^\dagger \mathscr{H},
    \qquad
    I_1 = \Tr \left( M^\dagger M \right),
    \qquad
    I_2 = \mathop{\mathrm{Re}} \mathop{\mathrm{Pf}} M,
    \qquad
    I_3 = \mathop{\mathrm{Im}} \mathop{\mathrm{Pf}} M,
    \qquad
    I_4 = \Tr \left[ \left( M^\dagger M \right)^2 \right],
\end{gather}
\end{small}
where $\mathop{\mathrm{Pf}} M$ is the Pfaffian of $M$ defined as
\begin{gather}\label{eq:Pfaffian}
    \mathop{\mathrm{Pf}} M = \frac1{2^2 2!} \varepsilon_{abcd} M_{ab} M_{cd} \quad \text{ for } n_F=2,
    \qquad
    \mathop{\mathrm{Pf}} M = \frac1{2^3 3!} \varepsilon_{abcdef} M_{ab} M_{cd} M_{ef}  \quad \text{ for } n_F=3,
\end{gather}
with $\varepsilon$ being the $2n_F$-dimensional Levi-Civita symbol ($\varepsilon_{12\dots (2n_F)}=+1$). In~the potential \eqref{eq:U}, $\lambda_{i2} = \lambda_{i3} = 0$ for all $i$ if $n_F = 3$; the invariant $I_3$ is CP odd, i.e., $\lambda_{3} = 0$ as well as $\lambda_{i3} = 0$ for $i=0$, 1, 2. Besides~we can always set $\lambda_{22} = 0$  because of the identity $I_1^2 -4 I_2^2 -4 I_3^2 -2 I_4=0$ that holds for $n_F = 2$. (For $n_F = 3$, the~corresponding term is nonrenormalizable and, thus, not taken into account.) One can derive and solve tadpole equations and diagonalize the quadratic forms in the scalar potential to obtain the mass spectrum of the (pseudo)scalar H-hadrons (see~\cite{hadronic}). Note that the term $I_0 I_1$ leads to a small mixing of the Higgs field and the singlet H-meson $\sigma'$ making the Higgs boson partially composite in this~model. 

If the hypercharges of H-quarks are set to zero, the~Lagrangian \eqref{eq:LQS1} is invariant under an additional symmetry---hyper $G$-parity~\cite{2010PhRvD..82k1701B,Antipin:2015xia}:
\begin{align}\label{eq:HGconjugation}
	Q^{\tilde{\text{G}}} = \varepsilon \omega Q^\text{C},
	\qquad
	S^{\tilde{\text{G}}} = \omega S^\text{C}.
\end{align}

Since H-gluons and all SM fields are left intact by \eqref{eq:HGconjugation}, the~lightest $\tilde G$-odd H-hadron becomes stable. It happens to be the neutral H-pion $\pi^0$.

Besides, the~numbers of doublet and singlet quarks are conserved in the model \eqref{eq:LQS1}, because~of two global U(1) symmetry groups of the Lagrangian. This makes two H-baryon states stable---the neutral singlet H-baryon $B$ and the lightest state in doublet $\mathscr{B}$, which carries a charge of $\pm 1/2$.

\subsection{Exotic States of New Colored~Objects}
\unskip
\subsubsection{Fractons}
Mixed states with nontrivial electroweak and dark QCD charges (or vice~versa, dark electroweak and ordinary QCD charges) can appear as fractionally charged colorless states (fractons), as first proposed in~\cite{fractonsMK}. It should be noted that the term ``fracton'' has appeared later in condensed matter physics~\cite{fractonCMP} defining the density of states on fractals. Here we use the original notion of fracton as fractionally charged colorless~state.

One can distinguish leptonic and hadronic X particles, forming correspondingly leptonic and hadronic~fractons. 

Leptonic fractons are originated from a new lepton like state X-lepton with fractional electromagnetic charge and dark QCD color. Dark QCD confinement binds it with dark QCD quarks in a dark colorless state, which possess fractional electromagnetic charge. Created in early universe X-leptons and their antiparticles are bound with corresponding dark QCD quarks and antiquarks to form leptonic fractons. Similar to the case of free quarks, studied in~\cite{ZOP}, negatively charged fractons can be bound with ordinary positively charged nuclei in stars thus protecting positively charged fractons from their annihilation. As a result, the amount of primordial fractons cannot decrease and as it is the case that free quarks~\cite{ZOP} exceed by several orders of magnitude experimental constraints in the search for fractionally charged particles in the terrestrial~matter.

Hadronic fractons appear when X-quarks, having dark electroweak and ordinary QCD charges bind with ordinary quarks in fractional charged colorless states. In~baryon asymmetric universe $\bar X$ antiquarks are bound with ordinary quarks $u$ in fractionally charged stable meson $\bar X u$, while X-quark forms fractionally charged $Xud$ baryon. If~dark electromagnetic attraction can overcome ordinary electromagnetic repulsion, in~the dense baryonic matter objects X-meson and X-baryon can recombine in charmonium-like $\bar X X$, decaying to ordinary particles, and~reduce the abundance of fractons below the experimental upper~limit.

\subsubsection{Fractionally Charged States in QCD-Like~Models}

As it follows from the above, in~the hypercolor SM extension with three doublets of additional H-quarks and in the case of zero hypercharges, the~Lagrangian contains interacting field of  hypermeson $\mathscr{B}$, the~lightest state in doublet, carrying fractional charges $\pm1/2$. These new H-mesons contain singlet H-quark, $s$. The~problem of such hyperparticle interpretation in $SU(6)$ extension is aggravated by the fact that the model invariance with respect to two $U(1)$ groups ensures of these objects stability (simultaneously with the one for neutral singlet H-baryon $B$). At~the same time, very strict restrictions are imposed on the concentration of fractionally charged particles in the modern universe. It can be said that within the framework of this $SU(6)$ scenario, such fractionally charged particles should either be created in some bound states (possibly, these are an analog of QCD tetraquarks), or~effectively~annihilated.

To describe their arising as some bound states from the very beginning, it is reasonable to try to rewrite the Lagrangian model in such a manner that compound H-quark objects could interact with other fields. In~other words, we should construct effective vertices for the H-tetraquark interactions with gauge bosons and other H-hadrons. Obviously, only electromagnetic interaction of these fractionally charged components cannot explain an appearance of such multi-H-quark states at the early stage of universe evolution. It means that, for this procedure, we need to analyze how hyperstrong interactions of H-quarks and H-gluons can work at very high temperatures and densities producing strongly connected systems instead of using effective $\sigma$-model construction. In~other words, we need to consider at high energy scale some hyper-QCD with analogous problems of the bound-state description in the framework of effective Lagrangian approach, with~an integration over some degrees of freedom or introducing of hyper-vacuum v.e.v.'s for the analysis of the bound states of H-quarks with sum rules method, for~example. Then, we would come to consideration of hyper-tetraquarks as H-quark bags, repeating  procedures and approaches of orthodox QCD at other scale. It means an investigation of the dynamics and magnitudes of H-quark and H-gluon vacuum condensates, providing an existence of H-quark “bags” with some mass, structure and specific interactions with fields of matter in a hot and dense universe.
So, this mechanism of H-strong interaction as the basis for the explaining of neutral H-quark states formation should be carefully considered in~detail.    

It can be, however, assumed that the symmetry breaking occurs and, correspondingly, fractionally charged objects appear at an early stage of evolution, apparently at the initial stage of inflation when their creation should be accompanied by a rapid transformation (annihilation) into ``ordinary'' stable neutral carriers of DM and fields of matter. Due to the presence of vertices describing their connection with W and Z-bosons in the $\sigma$-model framework, there is both the creation of (still massless) leptons and standard quarks, as~well as the transition of fractionally charged particles into neutral components of the DM. Possible quick disappearing of charged stable particles in annihilation reactions can be suggested as an alternative way to save the scenario with $U(1)$ accident~symmetries. 

It also should be suggested, when the annihilation can be---before or after the hyper-QCD symmetry breaking? In other words, when all masses are zero or when already massive fractionally charged particles transform into the set of H-neutral mesons and ordinary quarks and leptons? Because~these variants of high-energy sector of H-fields remains unstudied yet, the~question whether some traces of the initial process exist in the modern observable universe, namely, creation and following transformation of fractionally charged stable particles does not have an exact~answer. 

If we suppose that the process of intensive annihilation into neutral particles and/or states with integer charges takes place mostly through H-strong channels where H-quarks interact via H-gluons, it should be before these exotic H-mesons will scatter across the universe due to inflation. Early possible initial inhomogeneities in distribution of this type of matter should effectively transform into neutral stable objects of the orthodox DM, and~in these regions of increased density (prototypes of clumps) also should be produced photons with energies $\sim M_{DM}$ and streams of neutrinos and ordinary leptons and mesons. Perhaps inflation will keep some fingerprints of active processes of creation of photons and neutrino radiation from initial high density domains. However, the long evolution of the universe after inflation will inevitably try to hide the total and exact history of DM clumps' origin accompanied with high energy photons and~neutrinos.

An important channel of $\mathscr{B}$-mesons connection with the SM world follows from \eqref{eq:TM} and~\eqref{eq:dpi}---these fractionally charged exotic stable states have EW interactions with $W$-bosons and also interact with $Z$-bosons and photons. It means that they participate in the standard set of reactions with ordinary matter, i.e.,~they annihilate or produced with the EW cross sections $\sim$($10^2$--$10^4$) \mbox{pb}. 

There is, however, some specific feature of the scenario---two $\mathscr{B}$-states with the same fractional charges can annihilate into one charged $W$-boson, which decays into quarks or leptons conserving the charge. Assuming an existence of some initial asymmetry between these exotic states, it can be supposed that this is a source of the following asymmetry between standard quarks and, consequently, between~baryons. Certainly, then we shift the baryon asymmetry origin to an early stage of evolution relating the symmetry breaking possibly with initial stage of inflation and processes of the hidden mass~generation.   

Besides, these stable objects (as the other stable particles in H-color model), if~they were to exist, can be produced at collider by decaying virtual vector bosons. These reactions should have small (electroweak) cross sections manifesting itself as events with large missed energy and/or in generation of hadronic jets with corresponding fractional charge associated with~H-quarks.

It can be concluded, in~the hypercolor extension, as, probably, in~any extension with additional (heavy) fermions, new types of stable particles may appear. They can arise as the QCD-like bound states of additional quarks (mesons, baryons or diquarks) or in the framework of the extended $\sigma$-model. Dynamics of these new objects is defined by the model gauge symmetry, but~spectra of their masses and  the scale of their manifestations are unclear in advance. An~example is the appearance of fractionally charged stable objects when we extend the hypercolor symmetry from $SU(4)$ to $SU(6)$. The~peculiarities of such scenarios need to be considered carefully by studying the burn out kinetics of these stable particles, scales of symmetry breaking induced by new types of vacuum condensates. The~H-color symmetry allows to analyze specific features of these extended scenarios predicting the specific and measurable signals of new~objects. 

\subsubsection{Multiple Charged States in QCD and QCD-Like~Models}
$\Delta$-like states of new stable heavy quarks $Q$ are stable and bound much stronger than~ordinary hadrons since their chromo-Coulomb binding energy $\alpha_c^2 m_Q$ exceeds the energy of confinement $\Lambda \sim 300 \MeV$ at $m_Q > 7.5 \GeV$ for QCD running constant $\alpha_c = 0.2$ \cite{13d}. If~such stable states are charged, they should avoid geochemical constraints on anomalous isotopes~\cite{hadronic}. It was first noticed in~\cite{I} that this problem can be solved in the case of stable quark of the 4th family, if~the $U$ quark is the lightest and thus most stable quark in this family and the  generation of baryon asymmetry simultaneously provides generation of excess of $\bar U$ quarks. $\Delta^{--}$-like ($\bar U \bar U \bar U$) can be effectively hidden in nuclear interacting dark atom bound with primordial helium, as~soon as it is formed in Big Bang~Nucleosynthesis.

Models with new nonabelian symmetry can predict much wider class of multiple charged stable particles. Such models can provide non-supersymmetric composite Higgs solution for the SM problems of divergent mass of Higgs boson and of the origin of the scale of electroweak symmetry breaking. This approach acquires special interest in the lack of positive results of the searches for supersymmetric particles at the LHC
(see, e.g.,~\cite{cht} for review and references).

The Walking TechniColor (WTC) model involves two techniquarks, $U$ and $D$  transforming
under the adjoint representation of a SU(2) technicolor gauge group~\cite{Sannino:2004qp,Hong:2004td,Dietrich:2005jn,Dietrich:2005wk,Gudnason:2006ug}. A~neutral techniquark--antiquark state is associated with the Higgs boson. The~accompanying prediction is the existence of bosonic technibaryons  $UU$, $UD$, $DD$, and~their 
antiparticles. Conservation of the technibaryon number $TB$ leads to stability of the lightest~technibaryon.

Electric charges of $UU$, $UD$ and $DD$ are  given in terms of an arbitrary real number $q$ as $q+1$, $q$, and~$q-1$, respectively~\cite{hadronic,4,KK}. Compensation of anomalies requires existence of
technileptons $\nu'$ and $\zeta$ with charges $(1-3q)/2$ and $(-1-3q)/2$, respectively. If~technilepton number $L'$ is conserved the lightest technilepton is~stable.

In the early universe sphaleron transitions provide equilibrium relationship between $TB$,  baryon number $B$, of~lepton number $L$, and~ $L'$.~Freezing out of these transitions results in a balance between the stable techniparticle excess and the observed baryon asymmetry
of the universe~\cite{hadronic,KK,KK2}. Stable negatively charged techniparticles are bound in nuclear interacting dark atoms with primordial~helium. 

In the case of $q=1$, stable double charged techniparticles are possible~\cite{hadronic,4,KK,KK2}. 

Another choice of parameter $q$ results in a possibility of multiple $-2n$ charged stable techniparticles for $n>1$.~Possible stable multiple charged techniparticles are marked bold in Table~\ref{tab:X}~\cite{hadronic}.

\begin{table}[H]
\caption{List of possible integer charged techniparticles. Candidates for even charged constituents of dark atoms are marked bold~\cite{hadronic}.}
\centering
{\begin{tabular}{ccccccccc}
\toprule
\boldmath{$q$} & \boldmath{$UU(q+1)$} &  \boldmath{$UD(q)$} &  \boldmath{$DD(q-1)$} &   \boldmath{$\nu' (\dfrac{1-3q}{2} )$}& \boldmath{ $\zeta (\dfrac{-1-3q}{2} )$}\\
\midrule
1 & \textbf{ 2}&  1 & 0 & $-1$ & \boldmath{$-2$} \\
3 &   \textbf{4}&  3 & \textbf{2} & \boldmath{$-4$} & $-5$\\
5 &   \textbf{6 }&   5 &  \textbf{4 }& $-7$ & \boldmath{$-8 $}\\
7 &  \textbf{ 8 }&  7 &  \textbf{ 6 }&\boldmath{ $-10$} & $-11$ \\
\bottomrule
\end{tabular} \label{tab:X} }
\end{table}
\unskip
\subsection{Strongly Interacting Dark Matter~Candidates}
\unskip
\subsubsection{Stable Heavy Quark~Hadrons}
New heavy quarks possess strong QCD-type interaction, so they can form at hadronization phase of evolution the coupled states---new heavy mesons, $(q\bar{Q})$, and~fermions, $(qqQ)$,
$(qQQ)$, $(QQQ)$. Classification and the main properties of these new hadrons were considered in~\cite{hadronic} for the case of up, $U$, and~down, $D$, type of new quark $Q$. As~was
noted earlier, the~lightest neutral meson appears in the scenario with new quark of up type. In~Table~\ref{tab3}, we represent the main quantum characteristics and quark content of new
heavy mesons and fermions which contain new quarks of up type, $U$.

\begin{table}[H]
\caption{Characteristics and quark content of new hadrons.} \centering
\begin{tabular}{cccc}
\toprule
$\boldsymbol{J^P}$    & $\boldsymbol{T}$          & \textbf{Isotopic Content}         & \textbf{Quark Content} \\
\midrule
$0^-$             & $\frac{1}{2}$          & $M=(M^0\,M^-)$                    & $M^0=\bar{U}u$,\, $M^-=\bar{U}d$  \\
$\frac{1}{2}$     & 1                      & $B_1=(B_1^{++}\,B_1^+\,B_1^0)$    & $B_1^{++}=Uuu,B_1^+=Uud,B_1^0=Udd$ \\
$\frac{1}{2}$     & $\frac{1}{2}$          & $B_2=(B^{++}_2\,B^+_2)$           & $B^{++}_2=UUu,B^+_2=UUd$ \\
$\frac{1}{2}$     & 0                      & $(B^{++}_3)$                      & $B^{++}_3=UUU$ \\
\bottomrule
\end{tabular}
\label{tab3}
\end{table}

In Table~\ref{tab3}, the~mesons $M^0$ and baryons $B^+_1,\,B^{++}_2,\,B^{++}_3$ are stable and the lightest of them, neutral meson $M^0$, can be proposed as the DM candidate. The~evolution
of these new heavy hadrons was briefly considered in Ref.~\cite{hadronic}, where the process of burning out of heavy baryons was analyzed. Here, we represent the main properties of new
heavy mesons, $M^0=(u\bar{U})$ and $M^-=(d\bar{U})$, which~can lead to the characteristic signals of the hadronic dark~matter.

We have determined the mass of new mesons from the data on DM relic density with the help of the following~equality:
\begin{equation}\label{2.3.1}
(\sigma(M)v_r )^{Mod} =(\sigma v_r )^{Exp}.
\end{equation}

In Equation~(\ref{2.3.1}), the~left part of equality is the model value of annihilation cross section and the right part follows from the data on the modern relic concentration of DM, $(\sigma v_r)^{Exp}=2\times 10^{-9}$\, GeV$^{-2}$.
 To~calculate the model cross section $\sigma(M)$, which is a function of the mass $M$ of meson, we take into account the fact that the freezing-out temperature
$T_{freez}\approx M/30$ for the case of heavy DM particles is much greater than the temperature of QCD phase transition, $T_{QCD}\approx 0.15$ GeV. So, there are no coupled hadron states
at freezing-out stage and the dynamics of this stage is defined by the process of annihilation of new quark-antiquark pairs, $Q\bar{Q}\to gg, q\bar{q}$, where $g$ is a gluon and $q$ is a
standard quark. The~total cross section of these strong channels of annihilation was calculated in the limit of massless quarks in the final states and presented in
Ref.~\cite{hadronic}:
\begin{equation}\label{2.3.2}
(\sigma(M) )^{Mod}=\sigma(Q\bar{Q}\to gg,q\bar{q})\approx \frac{44\pi}{9}\frac{\alpha^2_s}{M^2}.
\end{equation}

From the expression (\ref{2.3.2}) and equality (\ref{2.3.1}) the estimation of the new quarks mass follows, \mbox{$M=10$ TeV,} which defines mass scale of new hadrons.~Electroweak channels of
annihilation, $Q\bar{Q}\to \gamma\gamma$, $ZZ$, $W^+W^-$, give small contribution into the total value of the annihilation cross section. It~is known that the Sommerfeld--Gamov--Sakharov (SGS)
enhancement effect can significantly modify the value of a cross section. Such enhancement takes place at hadronization stage of evolution, when SGS effect is caused by the light meson
exchange~\cite{hadronic}. At~quark-gluon stage this effect can be caused by $\gamma$ and $Z$-boson exchange only. As~was shown by numerical calculations~\cite{Cirelli07}, in~this case the
coefficient of enhancement is of the order of unity, i.e., the effect is~small.

  The value of mass-splitting in the doublet of neutral, $M^0=(u\bar{U})$, and~charged, $M^-=(d\bar{U})$, new heavy mesons plays an important role in HaDM description. We define
the value of mass-splitting as~follows:
\begin{equation}\label{2.3.3}
\Delta m=m(M^-)-m(M^0).
\end{equation}

In the case of standard heavy-light mesons the value $\Delta m$ is of the order of MeV, besides~this value is positive for the case of D-meson (up type heavy quark) and negative for the
case of K- and B-mesons (down type of heavy quark). New heavy mesons $M^0$ and $M^-$ are just the case of heavy-light meson, $m_Q\gg m_q$. From~the heavy quark symmetry, a direct analogy with standard heavy-light mesons follows, so, in~the case under consideration, we can assume that $\Delta m$ is positive and $\Delta m\sim$ MeV. Moreover, the~condition of
instability of the charged meson $M^-$ leads to inequality $\Delta m>m_e$, where~$m_e$ is the mass of electron. So, the charged partner of neutral DM particle has a unique decay channel with
very small phase space in a final state, $M^-\to M^0 e^-\bar{\nu}_e$. The~expression for the width of the charged meson is as follows~\cite{hadronic}:
\begin{equation}\label{2.3.4}
\Gamma(M^-)=\frac{G^2_F}{60\pi^3}|U_{ud}|^2(\Delta m^5-m^5_e),
\end{equation}
where $U_{ud}$ is the element of CKM matrix, which defines the charged transition $d\to u W$. From~the expression (\ref{2.3.4}) one can see that at $\Delta m\to m_e$, the~value of width
$\Gamma(M^-) \to 0$, i.e., the lifetime can be arbitrary large. For~instance, at~$\Delta m\sim 1$ MeV the lifetime $\tau \sim 10^5$ s. Thus, in~the scenario with hadronic DM, new
neutral meson $M^0$, as~a DM candidate, has charged metastable partner with the same mass, which should be taken into account in the process of co-annihilation. New heavy charged meson
appears in the process of collision of DM with ordinary matter, namely with leptons or nucleons. The~cross sections of these processes were calculated in Ref.~\cite{Kuksa20}, where the
main signals of hadronic DM manifestation are~considered.

One more principal feature of hadronic DM scenario is the effect of hyperfine splitting of excited states of new heavy hadrons. First of all, we should note that, in~contrast to fine
splitting, which is caused by change of quark content ($d\to u$) and has the value of the order of MeV, hyperfine splitting takes place for the states with the same quark content and has
much smaller value (of the order of KeV).

Further, we describe the effect of hyperfine splitting of ground and excited states, $\delta M_q=m(M^*_q)-m(M_q)$, where $M^*_q$ is an excited state of heavy meson $M$. Here, we consider
the simplest case of the lowest excited states of the new mesons $M_q=(q\bar{U})$. In~a direct analogy with the standard heavy-light (HL) mesons, $D_q=(cq)$ and $B_q=(\bar{b}q)$, we
define the ground and excited states in the terms $S^1_0$ and $S^1_1$ (the~classification with quantum numbers $L^{2s+1}_J$), or~$\frac{1}{2}(0^-)$ and $\frac{1}{2}(1^+)$ (the~classification $I(J^P)$). Here, $L$, $s$, $J$, $I$ and $P=(-1)^{1+L}$ are orbital momentum, spin, total momentum of the system, isospin and parity. We designate the~ground states $\frac{1}{2}(0^-)$
of the HL mesons as $D_q$, $B_q$ and $M_q$, while the excited states---as $D^*_q$, $B^*_q$ and $M^*_q$. We evaluate the mass-splitting between the excited and ground states,
$M^*_q$ and $M_q$, in analogy with standard splitting mechanism. This possibility is provided by the heavy quark symmetry which is the basic assumption of heavy quark
effective theory (HQET). Heavy~quark symmetry leads to the relations between the masses of excited states of $B$ and $D$ mesons~\cite{Ebert98}:
\begin{equation}\label{2.3.5}
m(B_2)-m(B_1)\approx\frac{m_c}{m_b}(m(D_2)-m(D_1)),
\end{equation}
where $m(B_k)$ and $m(D_k)$ are masses of $B_k$ and $D_k$, $m_c$ and $m_b$ are masses of constituent quarks. This~expression successfully describes the relation of splitting between
the lowest excited $\frac{1}{2}(1^-)$ and ground states $\frac{1}{2}(0^-)$ of $B$ and $D$ mesons:
\begin{equation}\label{2.3.6}
\frac{m(B^*)-m(B)}{m(D^*)-m(D)}\approx\frac{m_c}{m_b} \longrightarrow 0.32\approx 0.32\,(0.28).
\end{equation}

In (\ref{2.3.6}), we used $m(B^*)-m(B)=45$ MeV and $m(D^*)-m(D)=142$ MeV (see~\cite{PDG18}), \mbox{$m_c=1.55$ GeV} and $m_b=4.88$ GeV~\cite{Ebert98}. The~value of the relation in parentheses, (0.28),
follows from the data $m_c=1.32$ GeV and $M_b=4.74$ GeV~\cite{Mutuk18}. In~order to evaluate the mass-splitting in the doublet of new mesons $M_q=(q\bar{U})$, we used the relation
(\ref{2.3.5}) and took into consideration the equality $m(U)\approx m(M_q)=M$. Using the value of mass $M=10$ TeV, we get:
\begin{equation}\label{2.3.7}
\frac{\delta m(M)}{\delta m(B)}=\frac{m(M^*)-m(M)}{m(B^*)-m(B)}\approx\frac{m_b}{M} \longrightarrow \delta m(M)\approx\delta m(B)\frac{m_b}{M} \approx 2\, \mbox{KeV}.
\end{equation}

Thus, we get very small mass-splitting (hyperfine splitting) $\delta m$, which is much less than the fine splitting, $\delta m \ll \Delta m$. This effect follows from the HQFT prediction and is
caused by very large mass of new hadrons, i.e., the hyperfine splitting is a specific property of hadronic~DM.

The excited state of hadronic DM particles can manifest itself in the processes of interaction of neutral meson $M^0$ with radiation. Transition to the first excited state of the meson
$M^0=(u\bar{U})$ can be realized through the absorption of photons in KeV range which corresponds to the wavelength $\lambda \sim 10^{-9}$ cm. If~we assume that the meson
$M^0=(u\bar{U})$ has the size of the order of nucleon radius, $R_M\sim 10^{-13}$ cm, then $R_M \ll \lambda_{trans}$ and interaction of $M^0$ with photons is described by the higher terms of
multipole expansion of the charge distribution in the composite system $(u\bar{U})$. So, the~cross section of $\gamma M^0$ scattering is small and these mesons can be interpreted as
dark matter particles. At~$\lambda_{trans}\ll R_M$, i.e., $E_{\gamma}\gg 10$ MeV, the~cross section of interaction $\gamma M^0$ becomes large and dark matter becomes not absolutely
``dark''.

 Low-energy interaction of new heavy hadrons with the standard leptons is described in spectator approach by the effective Lagrangian of $WMM$ interaction in the differential
form~\cite{Kuksa20}:
\begin{equation}\label{2.3.8}
L^{eff}(WMM)=i G_{WM} U_{ik} W^{+\mu}(\bar{M}_{ui}\partial_{\mu} M_{dk} -\partial_{\mu}\bar{M}_{ui} M_{dk}) + h.c.,
\end{equation}
where $ui=u,c,t$; $dk=d,s,b$; $U_{ik}$ is corresponding element of CKM matrix, $M_{ui}=(ui\bar{U})$, $M_{dk}=(dk\bar{U})$, and~$G_{WM}=gU_{ud}/2\sqrt{2}$. The~value of effective
coupling constant $G_{WM}$ is equal to the fundamental constant in $W$-boson interaction with quark. Thus, the~spectator approach, which directly follows from the structure of process
at the fundamental (quark) level, is valid for the case of low-energy~interactions.

The structure of low-energy Lagrangian of Z-boson interaction $L^{eff}(ZMM)$ can be represented in analogy with $L^{eff}(WMM)$ by the simplest differential expression with regard to
the preservation of flavor ($qi\to qi$). In~contrast to (\ref{2.3.8}), effective coupling $G_{ZM}$ is caused by interactions of $Z$ with quarks $Q$ and $q$. So, in~this case, we meet
the problem of an effective coupling~definition.

Inelastic scattering of the low-energy leptons on the new heavy $M$ particles is defined by the $t$-channel diagram with $W$-boson in the intermediate state. In~the limit of zero
lepton mass and mass-splitting $\Delta M$ we get the cross section in the form:
\begin{equation}\label{2.3.9}
\sigma(l^- M^0\to \nu_l M^-)=\frac{3g^4 |U_{ud}|^2}{2^{10}\pi M^4_W}s(1-\frac{M^2}{s})^2,
\end{equation}
where $\sqrt{s}$ is full energy in the CMS. In~the non-relativistic case, expression (\ref{2.3.9}) can be represented in the form:
\begin{equation}\label{2.3.10}
\sigma(l^- M^0\to \nu_l M^-)=\frac{3G^2_F |U_{ud}|^2}{8\pi}(E_l+W)^2,
\end{equation}
where $E_l$ is the energy of lepton and $W=Mv^2/2$ is kinetic energy of the non-relativistic $M$-particle. The~process of lepton scattering on $M^0$ taking into account  final states is as
follows: $l^- M^0\to \nu_l M^-\to \nu_l M^0 e^-\bar{\nu}_e$. So, in~this process, the neutrino with energy $E_{\nu}\sim E_l$ appears together with $e^-\bar{\nu}_e$-pair (with~total energy
$E\sim \delta M$). For~the cross section of the process $\nu_l M^0 \to l^- M^+$ we get the same expression due to neglecting lepton mass in Equation~(\ref{2.3.9}).

 Heavy hadronic DM particles at the modern stage of evolution are non-relativistic, they have an average velocity $\sim 10^{-3}$ with respect to the galaxy. From~the kinematics of
the heavy DM particles and nucleon collisions it follows (see Ref.~\cite{hadronic}) that low-energy interaction can be described by the effective meson-exchange approach. The~nucleon-meson interaction was considered in~\cite{Vereshkov91} on the basis of the gauge scheme realization of symmetry $U(1)\times SU(3)$. This scheme was developed and applied to the
interaction of new heavy mesons with ordinary vector mesons~\cite{Bazhutov17,Beylin20}. The~part of physical Lagrangian which describes the interaction of nucleons and M-mesons with
ordinary vector mesons consists of two terms:
\begin{equation}\label{2.3.11}
L_{NMV}=L_{NV} + L_{MV}.
\end{equation}

In Equation~(\ref{2.3.11}) the first term describes interaction of nucleon with usual mesons:
\begin{align}\label{2.3.12}
L_{NV}&=g_{\omega} \omega_{\mu}(\bar{p}\gamma^{\mu}p+\bar{n}\gamma^{\mu}n) + \frac{1}{2} g\rho^0_{\mu}(\bar{p}\gamma^{\mu}p-\bar{n}\gamma^{\mu}n)\notag\\
      &+\frac{1}{\sqrt{2}} g\rho^+_{\mu}\bar{p}\gamma^{\mu}n+\frac{1}{\sqrt{2}} g\rho^-_{\mu}\bar{n}\gamma^{\mu}p,
\end{align}
where $g_{\omega}=\sqrt{3}g/2\sin \theta$, $g^2/4\pi \approx 3.4$ and $\sin \theta\approx 0.78$.

The second term in Equation~(\ref{2.3.11}) describes the interaction of $M$ particles with ordinary vector~mesons:
\begin{align}\label{2.3.13}
L_{MV}&=iG_{\omega M}\omega^{\mu}(\bar{M}^0 M^0_{,\mu}-\bar{M}^0_{,\mu} M^0 +M^+_{,\mu}M^--M^+ M^-_{,\mu})\notag\\
      &+\frac{ig}{2}\rho^0_{\mu} (\bar{M}^0 M^0_{,\mu}-\bar{M}^0_{,\mu} M^0 +M^+_{,\mu}M^--M^+ M^-_{,\mu})\notag\\
      &+\frac{ig}{\sqrt{2}} \rho^{+\mu}(\bar{M}^0 M^-_{,\mu}-\bar{M}^0_{,\mu} M^-) + \frac{ig}{\sqrt{2}} \rho^{-\mu}(M^+ M^0_{,\mu}-M^+_{,\mu} M^0).
\end{align}

In Equation~(\ref{2.3.13}), the~coupling constant $G_{\omega M}=g_{\omega}/3$. In~Ref.~\cite{hadronic}, it was shown that scalar mesons give very small contribution into $NM$ interaction and
we omit it here. Note, the~interactions of new mesons with ordinary pseudoscalar mesons (for instance, $\pi$-mesons) are absent due to parity conservation. This is an important
property which differs new heavy hadrons from the~nucleons.

Low-energy scattering of nucleons on new mesons is described by $t$-channel diagrams with ordinary vector and scalar mesons in the intermediate states. The~diagrams with pseudoscalar
mesons in the intermediate states are absent at the tree level, while the contribution of scalar mesons is negligible. So, the~dominant contribution is given by the vector-meson exchange (the vector
mesons being $\omega$ and $\rho$ mesons).

Now, we consider the kinematics of elastic scattering process $MN\to MN$, where $M=(M^0, M^-)$ and $N=(p,n)$. For~the case of non-relativistic particles, the~maximal value of momentum
transfer $Q^2=-q^2$ is $Q^2_{max}=(p k)^2\approx 4m^2_N v^2_r$. So, $Q_{max}\approx m_N v_r \sim 10^{-3}m_N$, the~value $Q_{max}$ is much less than the mass of the vector mesons $m_v$ ($m_v\sim
m_N$) and the meson-exchange model is~relevant.

 Using the expressions for the vertices from the expressions (\ref{2.3.12}) and (\ref{2.3.13}) we calculated the cross section of the process $N_a M_b\to N_a M_b$:
\begin{equation}\label{2.3.14}
\sigma(N_a M_b\to N_a M_b) = \frac{g^4 m^2_p}{16\pi m^4_v}(1+\frac{k_{ab}}{\sin^2\theta})^2,
\end{equation}
where $N_a=(p,n)$, $M_b=(M^0,M^-)$, $g^2/4\pi\approx 3.4$, $\sin\theta=1/\sqrt{3}$ and $k_{ab}=\pm 1$ for the case of proton, $p$, and~neutron, $n$. Equation~(\ref{2.3.14}) implies a rather large cross section, for~example $\sigma(pM^0\to pM^0)\approx 0.9$ barn. We should note that large cross section of $NM$-scattering can stipulate large interaction of
dark matter halo and galaxy at some stage of their evolution. The~problem of interaction between galaxies and dark matter halo was considered in details in Ref.~\cite{Wechsler18}.
Analysis of the low-energy elastic scattering $N_a M_b \to N_a M_b$ reveals an important peculiarity of the $NM$-interaction. Using the connection of potential and amplitude in Born
approximation we show that the potential of M-nucleon interaction at large distances ($d\sim m^{-1}_{\rho}$) has repulsive character~\cite{Bazhutov17,hadronic}. So, new heavy hadrons as
DM particles do not form coupled states with nucleon at low energy, i.e., at the modern stage of evolution. This effect makes it possible to escape the problem of anomalous hydrogen
and helium~\cite{hadronic}.

The processes of non-elastic scattering of type $N_a M_b \to N_c M_d$, where $N_a=(p,n)$ and $M_b=(M^0,M^-)$, have kinematics which is an analog of elastic scattering kinematics. In~this case, the~dominant contribution is caused by $t$-channel diagram with charged $\rho^{\pm}$-meson in the intermediate state. The~structure of expression for the cross section
explicitly describes the presence of threshold:
\begin{equation}\label{2.3.15}
\sigma(N_a M_b\to N_c M_d)=\frac{g^4 m}{8\pi v_r m^4_v}\sqrt{2m}[E_a-\Delta_{ab}]^{1/2},
\end{equation}
where $E_a\approx m_a v^2_r/2$, $m(N_a)=m_a\approx m_b\approx m$, $\Delta_{ab}$ is some combination of mass-splitting $\Delta M=m(M^+) - m(M^0)$ and $\Delta m=m_n - m_p \approx 1.4$
MeV, which depends on the structure of the initial and final states (see Table~2). Expression (\ref{2.3.15}) can be represented in another form:
\begin{equation}\label{2.3.16}
\sigma(N_a M_b\to N_c M_d)=\frac{g^4 m^2}{8\pi m^4_v}[1-\frac{\Delta_{ab}}{E_p}]^{1/2}.
\end{equation}

From (\ref{2.3.15}) it can be seen that the process of scattering has a threshold $E^{thr}_p = \Delta_{ab}$ when $\Delta_{ab}>0$.

In Table~\ref{tab4}, we present the expressions for the threshold in the case of basic reactions, namely~$pM^0\to nM^+$, $nM^+\to pM^0$, $nM^0\to pM^-$ and $pM^-\to nM^0$.

\begin{table}[H]
\caption{The threshold parameters $\Delta_{ab}$.} \centering
\begin{tabular}{ccc}
\toprule
$\boldsymbol{N_a M_b\to N_c M_d}$    & $\boldsymbol{\Delta_{ab}=f(\Delta M,\Delta m)}$  & \textbf{Signum $\boldsymbol{\Delta_{ab}}$}\\
\midrule
$pM^0\to nM^+$     & $\Delta_{p0}=\Delta M +\Delta m$          & $\Delta_{p0}>0$ (threshold)\\
$nM^+\to pM^0$     & $\Delta_{n+}=-\Delta M -\Delta m$          & $\Delta_{n+}<0$ (non-threshold)\\
$nM^0\to pM^-$     & $\Delta_{n0}=\Delta M -\Delta m$          & $\Delta_{n0} > 0\,\, (\Delta M > \Delta m)$\\
$pM^-\to nM^0$     & $\Delta_{p-}=-\Delta M +\Delta m$         & $\Delta_{p-} > 0\,\, (\Delta M < \Delta m)$\\
\bottomrule
\end{tabular}
\label{tab4}
\end{table}

Consider, for~example, the~first reaction, $pM^0\to nM^+$, where $E_p\approx m_p v^2_r/2$. The expression for threshold $E^{thr}_p=\Delta M +\Delta m\equiv \Delta_{p0}$ gives
the value of corresponding relative velocity $v^{thr}_r =\sqrt{2\Delta_{p0}/m_p}$. For~the case $\Delta_{p0}=10$ MeV we get the value of velocity $v^{thr}_r=0.1$ which is significantly
greater than the DM velocity now, $v_r\sim 10^{-3}$. So, this reaction is kinematically forbidden at the modern stage of evolution. The~third and fourth processes can be both
threshold and non-threshold depending on the value $\Delta M/\Delta m$. The~first, third and fourth processes lead to the intermediate (final) states with unstable particles. These
reactions go through two stages, for~example, $pM^0\to nM^+\to pe^-\bar{\nu}_e M^0 e^-\bar{\nu}_e$ and $nM^0\to pM^-\to pM^0e^-\bar{\nu}_e$. We should note that the reaction $nM^0\to
pM^-\to pM^0e^-\bar{\nu}_e$ is the most interesting due to presence of long-lived charge particle $M^-$. Note, the~indirect evidences of these particles were reported in
Ref.~\cite{Bazhutov17} (and references therein).  Thus, we get an interesting phenomenology of low-energy nucleon-DM scattering which has a specific~signature.

The process of annihilation $M^0\bar{M}^0\to X$, where a standard light particle appears in the final state $X$, has some peculiarities in HDM scenario. DM particles $M^0$, in~this
scenario, are composite and annihilation proceeds through both strong and EW channels. Note, the~theory of high-energy interaction of $M$ particles is unknown, however this reaction at
the sub-process level, $Q\bar{Q} \to q\bar{q},\,gg \to \mbox{jets}$, can~be approximately described. With~the help of this approach, we estimated the value of strong part of annihilation
cross section, which is described by the formula (\ref{2.3.2}). Here, we note that the dominant products of annihilation are the pairs of stable particles, $p\bar{p}$ and small
fraction of $e^+e^-,\,\nu\bar{\nu},\,2\gamma$ with total energy $E_{tot}\approx 2 M$.

In this subsection, we considered the principal phenomenological consequences of the hadronic composite DM---low-energy strong interaction, fine and hyperfine splitting of excited
states. Strong~interaction of DM with ordinary matter raises the problem of the connection between galaxies and their DM halos, which can contribute additional information to the modern
understanding of galaxy formation. Fine and hyperfine splitting in the set of new heavy mesons leads to
 the presence of metastable charged hadron and luminosity of hadronic DM. The~generation and possibility of registration of
heavy charged hadron in cosmic rays was briefly described. We also noted that the effect of hyperfine splitting generates the processes of electromagnetic transition and
recombination. These processes can be launched by the interactions of new hadrons with the ordinary matter and cosmic rays and lead to the effect of hadronic DM luminosity. Note, the~problems of hyperfine splitting and luminous DM become pressing now in view of the results of underground experiment XENON1T.

\subsubsection{Dark Atoms with Primordial~Helium}
Natural choice of parameters of sphaleron transitions between baryons, leptons and stable techniparticles leads in WTC model to balance between baryon symmetry and the excess of stable $-2n$ charged techniparticles, corresponding to the observed dark matter density at the mass of these particles in the TeV~range. 

For the sequential 4th family with electroweak SU(2) charges, similar balance can be established between the excess of stable $\bar U$ quarks and baryon~asymmetry.

Stable techniparticles behave like charged multiple charged leptons. ($\bar U \bar U \bar U$) states are also lepton-like, since their hadronic interaction is strongly suppressed~\cite{hadronic,13d}.

Excessive negatively even charged particles bind with primordial helium as soon as it is formed in the Big Bang~Nucleosynthesis. 

Double charged particles form $O$He dark atoms---a very nontrivial Bohr like atomic system with heavy lepton-like core and nuclear interacting helium shell with Bohr radius nearly equal to the size of helium~nucleus. 

$-2n$ charged techniparticles bind with $n$ primordial helium nuclei in Thomson like $X$He atoms with heavy lepton-like particle inside a nuclear droplet of $n$ helium~nuclei.

The Dark atom model has an advantage to explain the puzzles of direct dark matter searches by annual modulations of their low energy binding with Na nuclei. Strongly interacting shell of dark atoms provides their slowing down in terrestrial matter making this form of dark matter elusive for strategy of WIMP searches, involving significant nuclear recoil. However, dark atom interaction with nuclei can provide a low energy binding and the corresponding effect experiences annual~modulations. 

This explanation~\cite{iopKhlopov} is based on the following
picture of $O$He interaction with nuclei. $O$He is a neutral atom in the ground state, perturbed  by the Coulomb and nuclear forces of the approaching nucleus. The~sign of $O$He polarization changes with the distance. At~larger distances, Stark-like effect takes place---the nuclear Coulomb force polarizes $O$He so that the nucleus is attracted by the induced dipole moment of $O$He, while as soon as the perturbation by the nuclear force starts to dominate, the~nucleus polarizes $O$He in the opposite way so that He is situated more closely to the nucleus, resulting in the repulsive effect of the helium shell of $O$He. Qualitatively, it leads to a shallow potential well with a low energy bound state in $O$He-Na system, while such a state does not exist for heavy nuclei like xenon. A~quantitative description of $O$He-nucleus interaction with self-consistent account for the effects of nuclear and Coulomb forces is crucial to prove this explanation and such a description can lead to nuclear physics of $O$He (or $X$He), which determines their physical and astrophysical~effects. 

\section{New Physics of Strong Interaction in the~Galaxy}\label{galaxy}
\vspace{-6pt}
\subsection{New Components of Cosmic~Rays}
\unskip
\subsubsection{UHECR Interaction with Dark~Matter}

Mutual transformations of SM particles and their bound states in numerous reactions governing by conservation laws and known dynamics are necessarily supplemented by interactions with DM objects in the universe. In~addition to the gravitational interaction, the~main role is played by electroweak physics, ensuring the annihilation of dark matter (WIMPs) into standard particles. So,~processes in which DM candidates disappear generating fluxes of (unstable) mesons and baryons, nuclei, leptons~and photons can also be induced by strong interaction. In~any case, macroscopic cross sections of such annihilation processes are proportional to the DM density squared and provide the main set of possible signals carrying information about the spatial distribution and dynamics of the hidden mass. Photons, charged leptons and neutrinos can be considered as the most important carriers of such information; their spectra are measured by space detectors and telescopes with ever increasing accuracy and completeness. Obviously, due to lack of hopeful results from collider experiments, our~constant efforts to study the DM characteristics in the recent time inevitably reduce to different ideas and suggestions on indirect searches of DM manifestations in astrophysical data~\cite{KHL1,Cirelli1,RST_Rev,Gaskins,Arcadi,Belotsky,Indirect}.

Actually, all final SM particles arise either directly in the processes of annihilation (or decay) of the DM particles or at the stage of secondary mesons ($\pi^0, \, \pi^{\pm}, K$ etc.) decays. The~search for such signals which unambiguously produced by the hidden mass objects has been going on for a long time, but~there are no reliably confirmed signals yet. It should be noted that the DM annihilation with the maximum cross section occurs in regions of increased DM density, i.e.,~in the central regions of the halo near active galaxy nuclei (AGN), the~process can be also amplified in the possible DM clumps~\cite{Clumps1,Clumps2,Clumps3,Clumps_KHL,Clumps_Cherenkov}. In~other regions of galaxies, the~efficiency of annihilation signals decreases due to the low density of~DM.

Does not considering the decaying super-heavy DM, besides~annihilation reactions, the~presence of hidden mass can manifest itself in reactions of high-energy particles scattering on the DM particles. Such high-energy fluxes of cosmic rays permeate the entire universe, their composition, in~addition to the main components---protons and nuclei of various elements---includes electrons, photons, neutrinos. Furthermore, aside from the question of how particles of such high energies are generated and distributed across the universe~\cite{CR1,CR2,CR3,CR4,CR5,CR6}, the~processes of quasi-elastic and inelastic scattering of ultra-high energy cosmic rays (UHECR) by DM particles can give important information on the hidden mass dynamics and its spatial structures. Despite the fact that in this case the macroscopic cross sections are proportional only to the first power of the DM density, the~low probability of such processes is partially compensated by the specificity of the signature of the final states, i.e.,~the unique nature of the signals~\cite{Scatt0,Scatt1,Scatt2,Scatt3,Scatt4,Scatt5,Scatt6}.

Obviously, intergalactic magnetic fields strongly affect the propagation of charged leptons; therefore, information on the sources, conditions, and~principles of UHECR generation is mainly contained in neutrino and photon differential fluxes. Thus, the~cosmic rays interactions with hidden mass particles should be considered as a useful additional tool for studying DM in the universe. Analysis of such processes should contain not only the cross sections of CR interactions with hidden mass objects calculated in various DM scenarios, but~also an assessment of how the energy distributions and composition of UHECR can be changed due to quasi-elastic or inelastic interactions with the DM~particles.

As a testing ground for evaluating and discussing the processes of UHECR interaction with the DM particles, we consider the above described scenario of the SM extension due to additional fermions---hyperquarks. As~we have seen, in~the minimal version of the extension ($SU(4)\to Sp(4)$ H-color model),  the~DM candidates are stable neutral hyperpion (its charged unstable partners in the triplet of hyperpions have a masses greater by $\approx$160\, \mbox{MeV} and the neutral stable diquark, $B^0$~(see~above, Section \ref{sec212}). Here we will assume the masses of these two different DM candidates are practically equal; the case of an asymmetric DM, when one of the components is heavier than the other, can also be analyzed~\cite{doi:10.1142/S0217751X17500427}. The~reason is the calculated mass difference for these components clearly depends on the scale of renormalization and can be made nonzero. The~dependence itself on the renormalization follows from different H-quark currents corresponding to these stable objects in the scenario. As~it results from numerical analysis, the~mass splitting between two components of the DM cannot be more than (10--15)\, \mbox{GeV} for reasonable values of renormalization scale, $\mu$ = (100--1000)\, \mbox{GeV}. It, however, means that one type of the DM components can effectively transform into another with simultaneous production of soft leptons, neutrino and diffuse photons (see also~\cite{Two-Comp} where the DM components masses are rather different).

The reason of this soft radiation is that the dominant decay channel of secondary charged hyperpions is the decay $\tilde\pi^{\pm} \to \tilde \pi^0 \pi^{\pm}$ with subsequent leptonic decay of ordinary $\pi^{\pm}$. Therefore, the~signature of such a process---creation and subsequent decay of a charged hyperpion---is the appearance of a (decaying) muon and a muonic (anti) neutrino. Appearance and decay of charged H-pions is a characteristic feature of the scenario where additional heavy fermions form new H-hadron states with possibilities of their transformations into each other. Higher unstable H-quark bound states are not considered here under the assumption that their masses are sufficiently large (some interesting results about spectra of masses in H-color extensions basing on the lattice calculations can be found in~\cite{Lat1,Lat2}). Specifics of H-color dynamics in this type of models, with~two DM components, is that one of the DM candidates interacts via standard gauge bosons with the SM particles, but~the other one uses for such interactions only scalar exchanges by partly composite Higgs boson and its more heavier partner, $\tilde\sigma$-meson.

Thus, the~scenario under consideration represents of possible types of hidden mass components in the SM extensions with additional fermions. Due to the presence of DM components that are different in their nature and origin makes it possible to analyze various channels of interaction of these DM components with cosmic~rays.

Note that in the pioneering work on studying the channels of leptons scattering on the dark matter~\cite{Scatt0,Scatt2}, supersymmetry (SUSY) scenario was used as the basis, and, accordingly, neutralinos were considered as obvious candidates for the DM particles. In~this case, the~main subject of analysis was the secondary photons emitted during the scattering process. It seems to us that reactions in which not only leptons, as~a (small) part of cosmic ray flux, but also neutrinos play an active role are very informative. Besides, we mean not only neutrino-initiated processes of quasi-elastic interaction, but~also analysis of secondary neutrinos and photons produced in such scattering~events.

\begin{figure}[H]
\centering
\includegraphics[width=.45\linewidth]{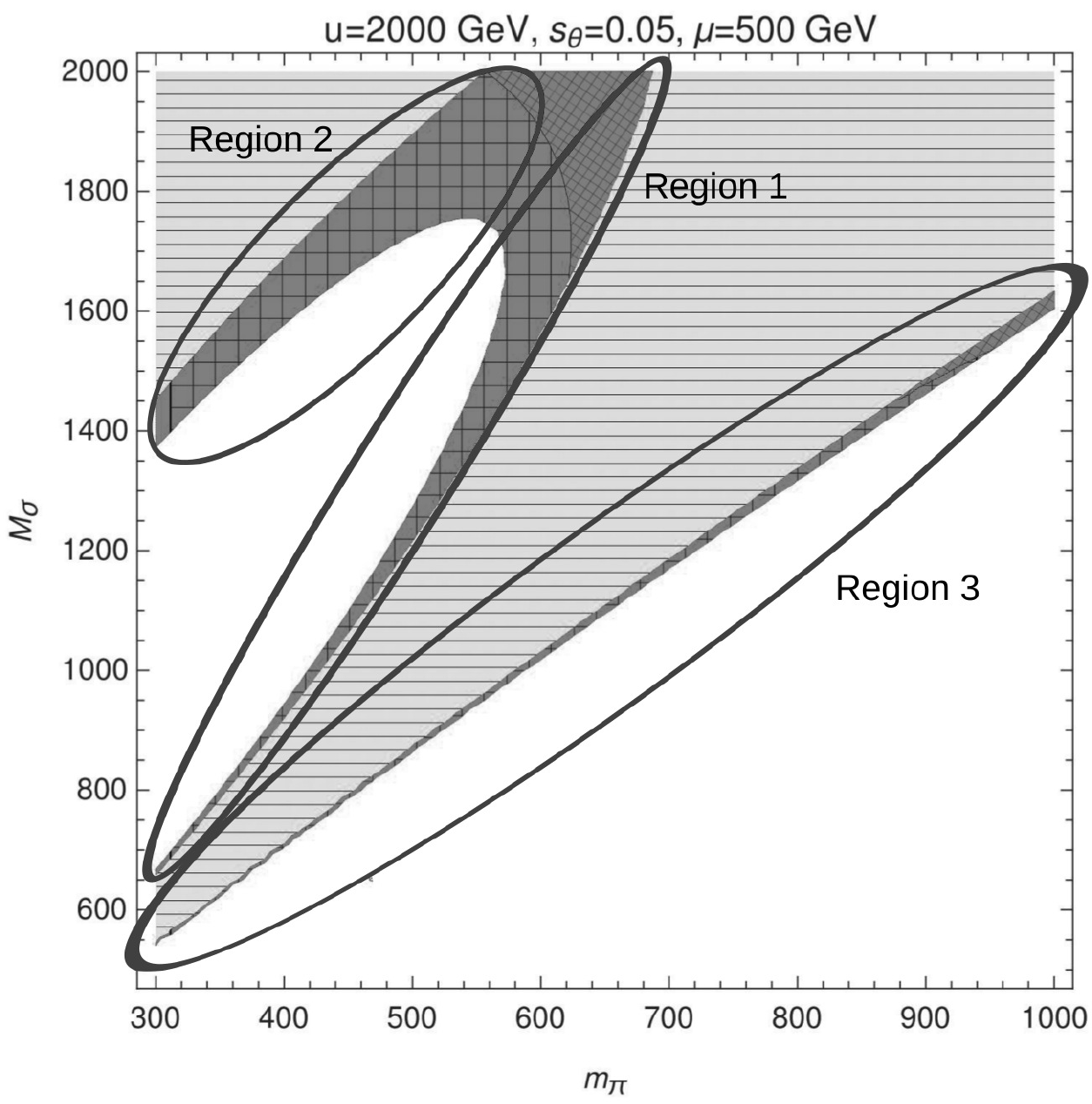}
\caption{Numerical solution of kinetic equations  system in a phase diagram in terms of $M_{\tilde \sigma}$ and $m_{\tilde \pi}$, other parameters are also~indicated.}
\label{fig1}
\end{figure}

\begin{figure}[H]
\centering
\includegraphics[width=.45\linewidth]{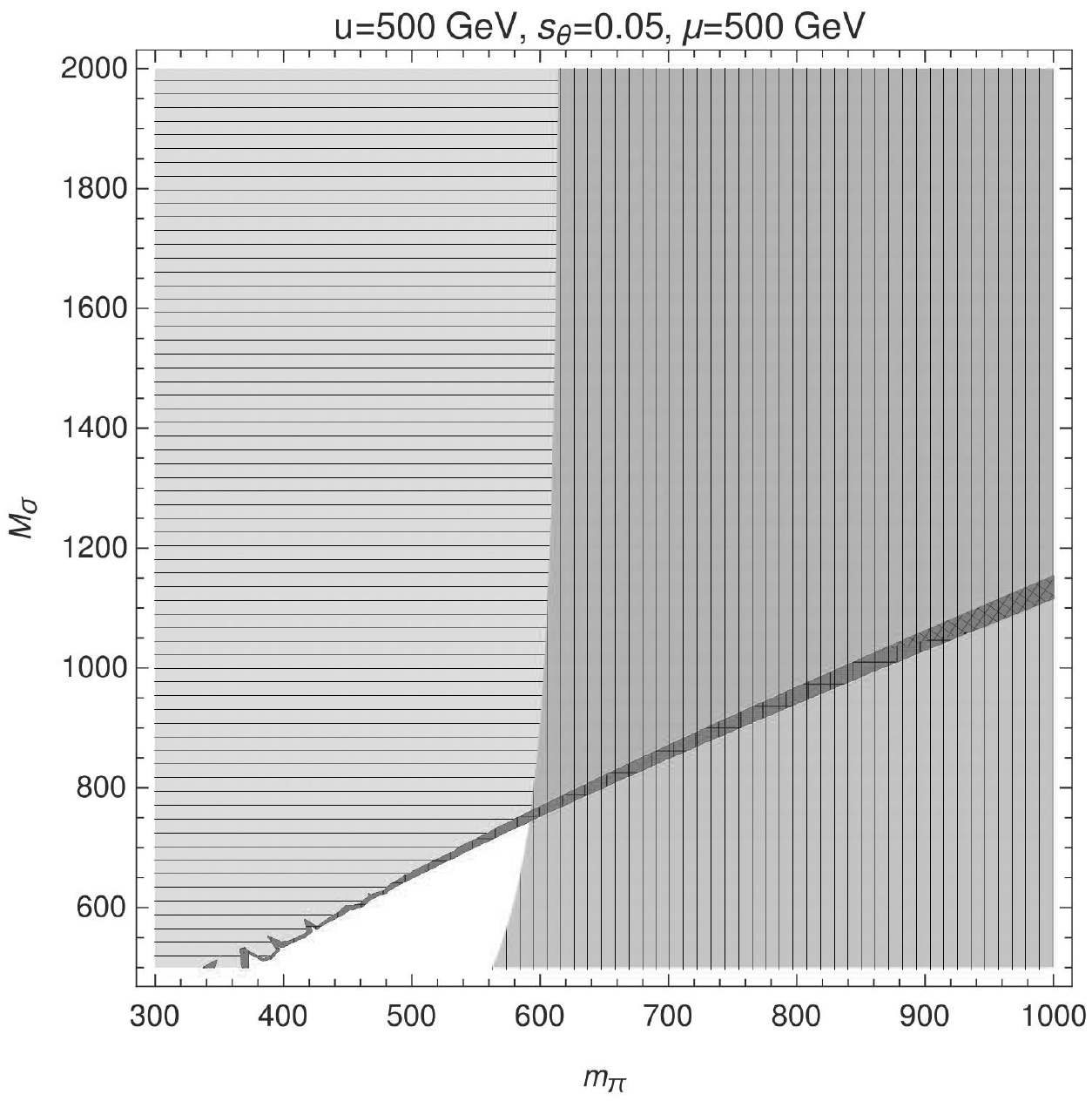}
\caption{Analogous phase diagram in terms of $M_{\tilde \sigma}$ and $m_{\tilde \pi}$, but~for much smaller $u$.}
\label{fig2}
\end{figure}

\begin{figure}[H]
\centering
\includegraphics[width=.5\linewidth]{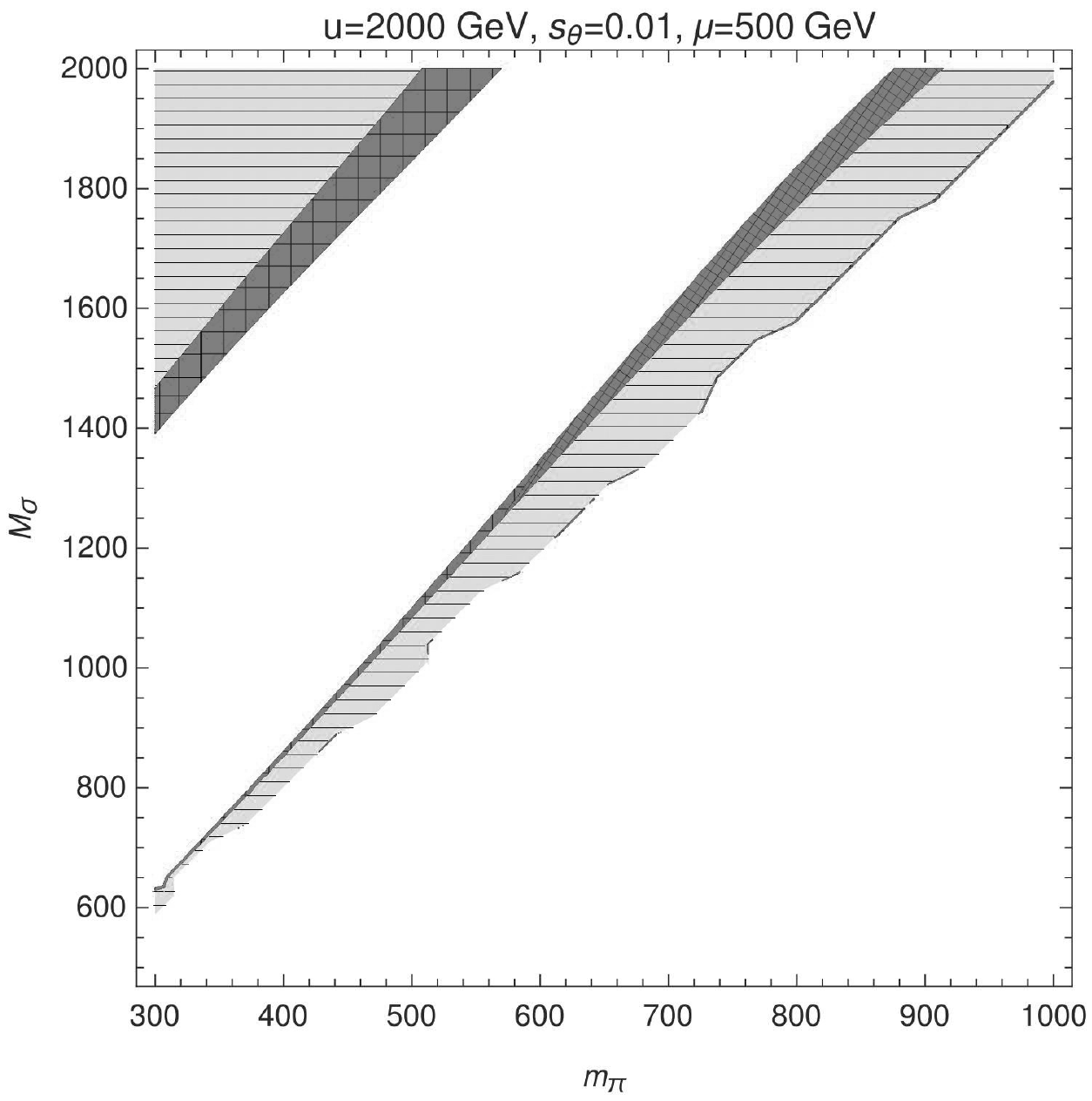}
\caption{Phase diagram in terms of $M_{\tilde \sigma}$ and $m_{\tilde \pi}$, the~same $u$ but the mixing is~smaller.}
\label{fig3}
\end{figure}

Masses of the DM components can be extracted from the system of five kinetic equations for the DM density assuming the existence of  freeze-out point for the annihilating DM candidates~\cite{Scatt5}.
Using~cross sections in all possible annihilation channels for the DM components, the~system of equations has been solved numerically. Then, we get some regions depending on the model parameters where the DM relic density is correct (in Figures~\ref{fig1}--\ref{fig3} these areas are crosshatched by vertical and horizontal lines) with H-pions fraction that is less than 25 percents: ($0.1047 \le \Omega h^2_{HP}+\Omega h^2_{HB} \le 0.1228$ and $\Omega h^2_{HP}/(\Omega h^2_{HP}+\Omega h^2_{HB}) \le 0.25$). The~crosshatching with oblique lines correspond to areas where all parameters are exactly the same, but~here H-pions fraction make up just over a quarter of the DM ($0.1047 \le \Omega h^2_{HP}+\Omega h^2_{HB} \le 0.1228$ and $0.25\le \Omega h^2_{HP}/(\Omega h^2_{HP}+\Omega h^2_{HB}) \le 0.4$). It is understood why $B$-component of DM dominates in all valid regions---this is because $B^0$-baryons interact with the world of ordinary particles only via H-quark and H-pion loops or through exchanges with scalar states, but~H-pions have tree level interactions with weak vector bosons, so they burn out much~faster.

Further, hatching with horizontal lines denotes regions where recent DM relic abundance is not explained by H-color candidates. Regions with vertical hatching are forbidden by underground-experiment data from XENON~collaboration.

Thus, results of kinetic analysis of two-component dark matter can be presented in the $M_{\tilde \sigma} - M_{\tilde \pi}$ plane  in Figure~\ref{fig1} as three allowed areas for the DM masses and  other model parameters (v.e.v. $u$, mixing angle $\theta$), namely:

{\bf Region 1}: $M_{\tilde{\sigma}}>2m_{\tilde \pi^0}$ and $u \ge M_{\tilde{\sigma}} $. At~small angles of mixing, $s_{\theta}$, and~large masses of H-pions it is possible to obtain a significant fraction of~H-pions. 

{\bf Region 2}: the same relation between $M_{\tilde{\sigma}}, \,\, m_{\tilde \pi^0},\,\, u$ but the H-pion mass is smaller, $m_{\tilde \pi}\approx$ 300--600~\mbox{GeV}, H-pion fraction is small~here.

{\bf Region 3}: $M_{\tilde{\sigma}}<2m_{\tilde \pi}$. This domain is always possible and it is presented in all figures. Note, decay $\tilde \sigma \to \tilde \pi \tilde \pi$ is prohibited. H-pion fraction in the DM relic can be large if the mass $m_{\tilde \pi^0}$ is large and the mixing angle is small. 
In Figures~\ref{fig2} and \ref{fig3}, we illustrate the regions changing for lower values of the v.e.v. $u$ and $\sin \theta$ .
\unskip

\unskip

Thus, knowing tree level cross sections of DM components annihilation, from~kinetics we estimate masses of stable hypercolor particles, $\tilde \pi^0$ and $B^0, \, \bar B^0$ in the range 0.6--1.2 \mbox{TeV}. Here we consider the case $\Delta M_{B \tilde \pi} \approx 0$. Note, due to connection between masses of H-pions and $\tilde \sigma$-meson for zero $h-\tilde\sigma$ mixing, mass of this scalar partner of ``nearly standard'' Higgs boson is also constrained in some range.
Now, with~the estimation of all masses in hand, we can analyze most simple process of quasi-elastic scattering $e^-\tilde\pi^0 \to \nu_e \tilde\pi^-$\cite{Scatt5} with a following decay of the charged~H-pion.

Cross section was calculated supposing the target (DM particle) gets a small portion of projectile energy, so $W$-boson in $t$-channel is close to its mass shell, the~momentum transfer is small and the final $\tilde \pi^0$ is not accelerated in this process (we return to discussion of this important possibility---acceleration of the DM objects up to TeV energies~\cite{Acc1,Acc2,Acc3,Acc4,Acc5,Acc6,Acc7}---somewhat later).  Then the secondary neutrino and leptons from the $W$ decay (in the channel $\tilde \pi^- \to \tilde \pi^0 W^- \to \tilde \pi^0e^-\bar \nu_e$) or from dominant channel of decay $\tilde \pi^- \to \tilde \pi^0 \pi^- \to \tilde \pi^0 \mu^-\bar \nu_{\mu}$ should have energies $\sim 10^2 \,\mbox{GeV}$ or even less (see Ref.~\cite{Beylin:2016kga} for details of the charged hyperpion decays). 
Then, the cross section of the process and also distributions on energy and angle of emission for secondary neutrino were found as and the number of possible neutrino events at IceCube produced by this reaction~\cite{Scatt5}.  Indeed, the~electron scattering on the DM objects can be interesting as a source of high-energy neutrino from $We\nu$ vertex or accelerated DM particles with masses $\sim 1 \, \mbox{TeV}$, when the momentum transferred is sufficiently~large. 

However, the~electron flux is only a small part of the cosmic ray total flux especially at energies $\geq 10^2 \, \mbox{TeV}$.~As~a result, we predict a very small fluxes of secondary neutrinos and, consequently, small~probability to detect such events at IceCube~\cite{Scatt5,Scatt6}.

At this moment, an~important feature of H-color model emerges---we have two DM components and the process of UHECR scattering should be considered for both types of neutral stable objects. Moreover, H-baryons $B^0$ are the largest part of the total DM amount, as~it follows from kinetic equations analysis. It follows from the absence of direct interaction of $B^0$-baryons with the standard gauge bosons and, consequently, with~the SM matter. So, the~burning out of this component is much slower than for $\tilde \pi^0$ particles.

It seems that there is a chance to introduce the $B^0$ interaction through H-pion and/or H-quark loops, however for the scattering channels these loops are exactly zero~\cite{Scatt5}. Thus, we need to consider more complex tree diagrams, in~particular, tree diagrams with the exchange of Higgs boson and its partner, $\tilde \sigma$-meson, in $t$-channel give dominant non-zero contribution to the process $e^-B \to \nu_e W^-B$. Virtual $W^-$-bosons eventually decay to $l \bar \nu_l$ or into light ordinary mesons. Of~course, there is similar scattering reaction with the scalar-state exchange, $e^-\tilde \pi^0 \to \nu_e W^-\tilde \pi^0$, whose amplitude is smaller by half as it is seen from the model Lagrangian. Here, we do not take into account small contributions from diagrams with H-quark loops, $hhZ$ and other multi-scalar vertices~\cite{Scatt6}. 

Note, diagrams of this type were recently considered and suggested for the analysis of neutrino scattering off nucleons~\cite{TRIDENT}, their significant contributions were confirmed by direct calculations. We,~however, found that these diagrams present dominant tree level part of cosmic particles scattering cross section off the DM. 
To calculate total width of the process with the final state $B^0e^-\nu \bar \nu$ or $\tilde \pi^0 e^-\nu \bar \nu$, we have used factorization method~\cite{Kuk_1,Kuk_2} considering independently amplitudes squared of subprocesses with intermediate $W$ and $Z$-bosons and then estimating the (negative) interference of these contributions. The~approach allows us to estimate with reasonable accuracy (no worse than $\sim$10\% due to approximate estimation of the interference) the cross section of an ``averaged'' process where the final electron and neutrinos are produced by different vertices, $W \to l \nu_l$ and $Z\to \nu_l \bar \nu_l$, which~practically coincide for the massless~leptons. 

So, without~using complex computer programs we get the value of total cross section and can estimate also the possibility to detect at IceCube the neutrino signal produced by the process of electron scattering off the DM. Again, for~these reactions we do not consider those phase space regions which correspond to acceleration of the initial DM particle (so called up-scattered dark matter).  In~other words, the final H-baryon (or neutral H-pion) is slow; nearly all energy of the incident electron is distributed between three final massless particles (electron and pair of neutrinos). Approximately, energies of secondary neutrinos are in the interval $\sim$$(E_e/3 - E_e)$.

 In calculations, we use two values of masses of H-baryon and $\tilde \sigma$-meson. Note that in the model there is a correlation between masses of $m_{\tilde \sigma}$ and $m_{\tilde \pi}$: $m^2_{\tilde \sigma} \approx 3\cdot m^2_{\tilde \pi}$.  This is an exact equality for zero mixing of Higgs boson and $\tilde \sigma$. Here, we suppose the splitting between masses of H-baryon and $\tilde \pi$ is very small. It was found that the cross section strongly depends on the mass of DM particle and grows with the mass~increasing.
 
Now, we should note an interesting type of the scattering processes produced by high-energy neutrino off the DM components. Namely, there is a small probability to find in the UHECR content electrons of very high energies, which will initiate creation of high-energy neutrino in the scattering. Nevertheless, we calculate cross section values for high energy region despite the fact that the cosmic electrons flux noticeably decreases for these energies; the hope is based on estimation of effective areas for IceCube---high-energy neutrinos can be detected with a larger probability. Unfortunately, cosmic electron motion is strongly affected by galactic magnetic fields, so their sources are hardly identified both in direction and in intensity. Certainly, if~we could  separate, in all experimental data, signals from high-energy electron scattering with specific set of final states, it would manifest itself on the target with defined properties; and, perhaps, it would have been the DM object. Alas, such signals are practically unobserved for the ground neutrino telescopes due to large and constant background from the Sun neutrino and neutrinos resulting from decays of mesons and baryons produced by cosmic rays interactions with nuclei in the Earth atmosphere. We also note that in processes with the production of secondary neutrinos, acceleration of DM particles can occur simultaneously; such processes can themselves be initiated by incident high-energy neutrinos~\cite{Neutrino_Review,Acc8Neutrino,Acc9Neutrino}. In~other words, reactions with participation of high-energy neutrinos which can be accompanied with the DM accelerated seem as informative and important, especially because both of these particles are, in~fact, messengers from regions of high DM density---regions near AGN or from possible DM inhomogeneities of some other nature---and early epoch of the universe~\cite{Neutrino_Early}.

In more detail, the~secondary neutrino fluxes calculated are very small in comparison with expected neutrino fluxes from AGN which can be $\sim 10^5$ cm$^{-2}$s$^{-1}$sr$^{-1}$.~Atmospheric neutrino fluxes with neutrino energies $\leq$2 TeV are also much larger~\cite{NeutrinoFlux1,NeutrinoFlux2,NeutrinoFluxAtm,Eff7Rev}: $\sim$10$^{-10}$--10$^{-9}$ cm$^{-2}$sr$^{-1}$s$^{-1}$. Namely, we get that the secondary neutrino flux resulted from the cosmic electrons scattering is $\sim$10$^{-19}$--10$^{-22}$~cm$^{-2}$sr$^{-1}$s$^{-1}$\cite{Scatt6}.
If, however, intergalactic neutrinos with very high energy come to the Earth (their sources can be jets from blazars), their scattering on the DM from halo can be marked by the specific reaction: 
$\nu_l+\mbox{DM} \to \bar \nu_l+Z^* +\mbox{DM} \to \bar \nu_l+ \nu_k\bar \nu_k+\mbox{DM}$. Of~course, a virtual $Z$-boson can decay into light mesons producing a number of soft leptons, neutrino and photons, but~some correlation in energies and directions between high-energy secondary neutrinos can be detected. Obviously, cross section of creation of secondary quarks or leptons by virtual $Z$-boson is resonantly amplified when the $Z$-boson is near its mass shell. In~any case, for~this process we evaluate cross section as $\sim$(20--300) \mbox{pb} for initial neutrino energies $\sim$(10--100) \mbox{TeV}.

Let us note some points which are important for study of the cosmic rays scattering off the DM. We suppose that, independently of the type of the SM extension, possibility of scalar exchange in the scattering channel results in a strong dependence of the cross section on the DM particle mass. It~was found in the H-color extension, namely, the~changing (increasing) of the DM component mass of 10\% leads to the cross section growth by up to 50\%. The~opening of channels with scalar exchanges allows us to consider new ways to produce secondary high-energy leptons, neutrino and photons by UHECR scattering off the DM. Besides, these reactions can result in acceleration of the DM particles despite that this process takes place in t-channel.  Now, resonant amplifying~\cite{Neutrino_Review} is possible only due to that part of total amplitude which describes producing of leptons or quark pairs in virtual gauge boson~decays.

\subsubsection{Creation of New Components in the UHECR~Sources}

Cosmic rays of ultrahigh energies have a low intensity which rapidly decreases with increasing energy; for example, at~the border of atmosphere, a~detector with an area of  1 m$^2$ can detect about 100~cosmic particles per year with $E\geq 10^3$ \mbox{TeV}. With~such small fluxes of UHECR, the~analysis of extended air showers (EAS) generated by high-energy particles in the Earth's atmosphere becomes an effective method of studying them. Composition of the initial radiation changes due to creation and decay of new particles that generate a nuclear-electromagnetic cascade. The~path traversed by particles in the atmosphere is much larger than the average range of inelastic interactions between protons and nuclei. Secondary particles forming an EAS, can be detected at large distances, up~to $10 ^3\, \mbox{m}$ or even more from the shower~axis. 

Standard picture of EAS production is based mostly on physics of high-energy protons (or light nuclei from cosmic rays) interaction with nuclei in atmosphere; then, secondary nucleons, pions, kaons, hyperons, leptons and photons are born.  In~the initial act of EAS generation leading secondary particle keeps about $\sim$50\% of its initial energy, so is able to interact several times in the atmosphere. At~high energies, $\geq$100 \mbox{TeV}, a~significant part of unstable secondary particles ($\pi$- and $K$-mesons) do not decay on the path of the order of one path of the inelastic nuclear interaction, they again interact with the nuclei, forming new decaying charged and neutral mesons which produce (decaying with lifetime \mbox{$\sim$2 $\times$ 10$^{-6}$ s}) muons, neutrinos and~photons. 

Of course, along with a shower of nuclear-active particles, it develops an electron--photon cascade in the atmosphere due to fast decays of neutral pions into two gammas. So, an~electron--photon component of the shower arises and evolves. Then, photons produce electron--positron pairs interacting with the medium, and~these charged leptons again give high-energy photons due to bremsstrahlung on the nuclei. Obviously, the~multiplication of particles in these showers is defined by energy dissipation processes for every type of interaction between mesons, nucleons and nuclei or mesons decays. These~interactions have been studied with all necessary detail.~The~passage of an EAS through the atmosphere is also accompanied by optical radiation: Cherenkov and ionization radiation. So, this~standard picture of the emergence and development of EAS generated by high-energy protons or nuclei must be supplemented since EAS can also be generated by high-energy neutrinos and, possibly, by~high-energy accelerated neutral DM particles~\cite{EAS_Rev,Acc10CR,AccDM}.

Neutrinos generate electromagnetic and hadronic showers due to deep inelastic (or quasi-elastic) scattering off nucleons in case of charged or neutral current interaction. Note that in~the last case there arise some special contributions providing interactions of $B^0$ component with standard fermions through scalar exchanges. These  diagrams, which are called as ``trident''~\cite{Scatt5,Scatt6,TRIDENT}), are especially important for the cosmic rays (electrons, neutrino, protons) interaction not only with nucleons but with the DM objects, in~particular, for~various multi-component DM models with scalar interaction of the DM with gauge bosons~currents.

It is important to note that the showers generated by neutrinos and nucleons can be effectively discriminated due to small cross section of neutrino interaction with nuclei in atmosphere (this~argument can also be used for separation of possible showers producing by high-energy neutral heavy DM particle~\cite{EUSO,Neutrino_Fargion}). As~a result, EAS induced by neutral particles begins its development deeply in atmosphere (where the density increases significantly) in comparison with EAS generated by cosmic protons. Produced by galactic neutrinos EAS are highly inclined, despite this, there is a set of parameters (Cherenkov light, depth of shower maximum in the atmosphere, duration of the shower in dependence on the altitude and, correspondingly, density) allowing to separate neutrino showers from proton ones~\cite{EAS_Rev,EUSO,ANTARES}. These characteristics of EAS can also be used to mark EAS which are produced by boosted DM~particles. 

Remind, the~possibility to accelerate light DM particles in the scattering of high-energy cosmic rays off the DM was supposed and numerically analyzed in~\cite{Acc1,Acc2,Acc3,Acc4,Acc5,Acc6,Acc7,Acc10CR,AccDM,AccCRDM}. This possibility to boost the DM was also confirmed for heavy DM objects with mass $\sim$1 \mbox{TeV}, more exactly, we have considered and approximately calculated scattering of protons of high energy, up~to $200 \,\mbox{TeV}$ on the DM particles from halo, in~the region of most large DM density---near AGN~\cite{Rep_Bled}. In~other words, we~consider interaction of protons from blazar’s jets with heavy DM particles. In~the framework of the H-color scenario, in~this interaction of protons with two DM components a significant part of protons energy can be transferred due to charged current to heavy H-pion and to both DM component in the trident type diagrams involving scalar~exchange.
  
For initial protons with energy $200\, \mbox{TeV}$ cross section of the scattering process where final charged H-pion is produced with energies (40--50) \mbox{TeV} is $\sim$(10--15) \mbox{pb}. This charged H-pion decays predominantly as $\tilde \pi^{\pm} \to \tilde \pi^o \pi^{\pm}$ with the width $\Gamma\to 3 \times10^{-15} \, \mbox{GeV}$. So, we get also secondary muon (which again decays) and muonic~antineutrino.

In fact, in~this deep inelastic reaction the main charged component of UHECR (protons) originated from blazar jet disappear transforming finally into flux of high energy neutrino and leptons. More~exactly, our estimations demonstrates that $\sim$(10--25)\% of the proton energy is transferred to heavy neutral component of the DM with cross section $\approx$(10--100) \mbox{pb}. In~the scattering channel described by “trident” diagrams total cross section is of the same order but there appear additional neutrinos, for~example, generated in the resonant decay of intermediate~Z-boson.  

Despite the cross section not being large, it is an example of accelerating even heavy DM particles up to significant energies. Most importantly, this boosted neutral particle, as~light neutrino, will pass away from the DM halo moving in the constant direction because it interacts with the matter very slowly.    
So, this rare process when the charge component of cosmic rays can be ruined in the deep inelastic reaction and as a result neutral DM particle moves like a neutrino towards the Earth. Remind~that above considered high-energy electron scattering off the DM can also accelerate the DM but in quasi-elastic process high energy neutrino are generated with more~probability. 

Thus, from~this brief description of some processes of scattering of high-energy cosmic ray particles off the DM we can conclude that these reactions can enrich the cosmic rays composition with boosted heavy neutral DM particles~\cite{CRComp}. At~energies of these projectiles $\sim$(10--100) \mbox{TeV} cross sections of their interactions with nucleons and  nuclei, $\sim$(10$^{-34}$--10$^{-37}$) cm$^2$, are compared with cross sections of neutrino-nucleons scattering. In~this deep-inelastic process nucleons or nuclei are transformed a multiparticle final states consisting of charged leptons, photons and neutrino. Additional neutrinos are generated by the charged H-pion decay and in the processes with resonant decay of Z-boson. So, the~accelerated neutral DM components can produce rare events - specific types of EAS. Certainly, these EAS in the atmosphere should be separated in some manner from other types of~showers. 

It is known that as usually cosmic rays generate a shower of secondary particles which are mainly muons, electrons and photons. They go to ground detectors and can be fixed as measured signals registering also due to fluorescence and Cherenkov light, and~radio emission generated by charged component, electrons, in~atmosphere of the Earth. It seems, such type of shower is similar to neutrino induced shower and its initial point also should be deeply in atmosphere, however, the~neutral DM particle can not disappear from the EAS composition and will interact with the ground detector producing some radiation from secondary electrons or from excited nuclei in the detector. The~DM showers, as~they generated by intergalactic DM objects which were accelerated by UHECR or AGN jets from halo of other galaxies, or~DM particles boosted  from halo of our galaxy by intergalactic UHECR do not have to be mostly inclined or nearly horizontal. It is supposed, these accelerated DM components and EAS produced by them should be distributed more or less isotropic. May be, the~EAS axis can be connected with direction to some blazar, as~it was found for some very high-energy neutrino events at~IceCube. 

So, we can conclude that EAS from heavy DM particles are distinguished from EAS generated by protons or neutrino because in the former  event the shower contains in his composition neutral stable object up to the final moment when this fast DM particle scattered on nucleon in the detector (see~also~\cite{DM_Search} and references therein). To the~contrary, in~composition of EAS which was induced by neutrino or protons (or light nuclei), there is no any heavy stable particles, only leptons, photons and neutrino are detected as final states. Note also that interaction of DM components with nucleons in detector should have specific signature: the scattering in charged current channel is accompanied with creation and following decay of charged H-pion, so, the~event can be seen due to charged lepton bremsstrahlung. We hope that observing and measuring the characteristics new types of EAS containing heavy  neutral stable particles will be possible at modern complex LHAASO~\cite{LHAASO}, in~other words, the~DM candidates can manifest itself in a specific types of~EAS.

Ionization of dark atoms by high energy cosmic rays or in supernova explosions can lead to formation of an anomalous cosmic ray flux of stable multiple charged leptons. The~search for such component and the possibility to discriminate the corresponding air-showers are challenges for LHAASO~experiments.

\subsection{Multimessenger Probes for New Physics~Effects}

As was noted above, UHECR can result in specific interactions with DM particles, which are concentrated in halo with the largest density near AGN, giving some special signatures. First, in~the scattering processes the target (slow heavy stable DM component) can receive a significant portion of energy of initial projectile. So, the~DM particle can be effectively accelerated mostly in the direction close to the projectile direction. However, in~the two-component DM scenario considered, if~the interaction is provided by charged current with the W-exchange neutral H-pion component transforms into charged H-pion decaying with generation of lepton and neutrinos from $\pi^{\pm}$ and muon decays. This charged H-pion, in~fact, decays in flight, so it is possible to estimate energies of final particles and their angles of emission. Indeed, these reactions are rare because of small density of DM, especially if we consider the intergalactic UHECR scattering off the DM halo of our~galaxy.     

Although secondary neutrinos (prompt neutrino from $Z\nu\bar \nu$ vertex, neutrino from $\tilde \pi^{\pm}$ decay and neutrinos from decaying mesons generated by initial proton) are emitted and scattered at different angles, there is non-zero probability to detect neutrino from this reaction at IceCube together with observation of specific shower associated with the DM component at LHAASO, for~example. Certainly, the~probability of such events which are separated by definite interval of time (its magnitude is defined by energies and velocities of the neutrino and DM particle) is evaluated as very small. However, if~such twin event would be detected, it could be an important manifestation of the UHECR interaction with the DM. More precisely, measuring the characteristics of such an event---the delay time between signals (registration of a high-energy neutrino and a shower of particles), the~energy release of the EAS, its composition, as~well as the establishment of the fact of interaction of neutral object (heavy DM particle) with the substance of the detector (for example, by~low-energy radiation)---could improve our understanding of the DM nature and the features of its interaction with neutrinos, leptons and nucleons. These probes can shed light on the possible hadronic and hadron-like particles of dark matter and their~spectroscopy.
\section{Conclusions}
Over recent decades, the mainstream of BSM physics has been concentrated on the search for direct and indirect effects of supersymmetry (SUSY). SUSY partners of SM particles with masses in the range of several hundred GeV--1 TeV were expected to be found at the LHC. The~lightest stable neutral SUSY particle could nicely implement the WIMP miracle and was considered as the preferable candidate the cosmological dark~matter.

However, SUSY particles were not found up to now at the LHC. The~results of direct underground WIMP searches are controversial and the positive result of DM searches by DAMA/NaI and DAMA/LIBRA experiments can hardly be interpreted in the terms of~WIMPs.

It is reasonable to extend the field of studies of the new physics and, in~particular, to~consider non-SUSY BSM models.~New stable color particles can provide candidates for dark matter with hadronic interaction, while new BSM nonabelian symmetry can increase the list of WIMP-like dark matter~candidates.

We have presented in this review various nontrivial features of new physics of strong interaction and their possible physical, astrophysical and cosmological signatures.
Experimental probes for these signature will shed new light on possible role of various forms of strong interaction in the dark~universe.

\vspace{6pt} 
\authorcontributions{Review by V.B., M.K., V.K. and N.V. The authors contributed equally to this~work. All authors have read and agreed to the published version of the manuscript.}

\funding{The work was supported by grant of the Russian Science Foundation (Project No-18-12-00213).
}

\acknowledgments{\textls[-15]{We express our gratitude to Andrea Addazi, Konstantin Belotsky, Maxim Bezuglov, Timur~Bikbaev, Vladimir Korchagin, Antonino Marciano, Andrey Mayorov and Egor Tretiakov for useful~discussions.}}

\conflictsofinterest{The authors declare no conflict of~interest.} 

\reftitle{References}

\end{document}